\documentclass[titlepage,twocolumn]{revtex4-1}

\usepackage{bbold}
\usepackage{amssymb}
\usepackage{graphics}
\usepackage{graphicx}
\usepackage{epstopdf}
\usepackage{multirow}
\usepackage{amsmath}
\usepackage{amsthm}
\usepackage{amstext}
\usepackage{amscd}
\usepackage{epsfig}
\usepackage[mathscr]{eucal}
\usepackage{times}
\usepackage{srcltx}
\usepackage{shadow}
\usepackage{url}
\usepackage{color}
\usepackage[all]{xy}

\usepackage[most]{tcolorbox}

\usepackage{hyperref}

\usepackage{textcomp}

\begin{document}

\title{Tissue failure propagation as mediated by circulatory flow}

\author{Gurdip Uppal${}^{1,*}$, Gokhan Bahcecioglu${}^{2,*}$, Pinar Zorlutuna${}^{2,**}$, Dervis Can Vural${}^{1,**}$}
\affiliation{${}^{1}$Department of Physics, University of Notre Dame, IN, USA\\${}^{2}\text{Department of Aerospace and Mechanical Engineering, University of Notre Dame, IN, USA}$\\${}^{*}$Equally contributing first authors\\${}^{**}$Equally contributing corresponding authors\\${}^{**}$Correspondence: pzorlutu@nd.edu, dvural@nd.edu }








\begin{abstract}
Aging is driven by subcellular processes that are relatively well-understood. However the qualitative mechanisms and quantitative dynamics of how these micro-level failures cascade to a macro-level catastrophe in a tissue or organs remain largely unexplored. Here we experimentally and theoretically study how cell failure propagates in a synthetic tissue in the presence of advective flow. We argue that cells secrete cooperative factors, thereby forming a network of interdependence governed by diffusion and flow, which fails with a propagating front parallel to advective circulation. 
\end{abstract}


\maketitle

\onecolumngrid
\begin{tcolorbox}[width=\textwidth]
Significance: Mortality rates typically increase for complex organisms as they age. This leads us to suggest that aging depends on interactions between cells. As cells become damaged, the effect propagates to other cells, eventually leading to a systemic catastrophe. Yet it is unclear how this failure dynamically propagates. Here we present experiments with synthetic tissues and analogous analytical models to investigate the dynamics of failure propagation. Our main contribution is a detailed investigation of failure propagation when interactions are mediated by advective flow. We find analytical expressions for when a pronounced propagation occurs, its velocity, and acceleration in terms of system parameters. 
\end{tcolorbox}
\twocolumngrid
\section*{Introduction}
The aging and death of an organism is typically attributed to subcellular mechanisms such as reactive oxygen species damage or slowing down of tissue repair due to shortening telomeres. However, an organism does not die because it gradually runs out of cells, but rather, cellular level failures cascade to tissues and organs that lead to a relatively sudden systemic catastrophe. While microscopic mechanisms of cellular malfunction are relatively well studied \cite{rattan2006theories,blackburn2000telomere,murabito2012search,shay2000hayflick}, how failure spreads from the subcellular level to tissues and organs and ultimately the organism is largely unknown.

In \cite{gavrilov2001reliability} the failure of an organism was modeled as a reliability circuit where cells within an organ are connected by OR gates, so that an organ fails when all its cells fail), and the organs are connected by AND gates, so that organism dies as soon as one organ fails. In \cite{vural2014aging}, aging was described as failures taking place on a complex network of interdependent building blocks. Here, when a node in the network malfunctions, so will those that depend on it. As a result, few random microscopic failures can propagate into many others, ultimately leading to a catastrophe. \cite{vural2014aging} could bridge micro scale malfunctions with their experimentally observed macroscopic manifestations such as organismic death and population demographics, and fit experimental data such as \cite{vaupel1998biodemographic} and \cite{caughley1966mortality} (also cf. appendix of \cite{stroustrup2016temporal}). The network models have also been insightful in studying frailty \cite{farrell2016network,mitnitski2017aging}.

While describing a complex organ or an entire organism as a random network of interdependencies is a useful starting point to understand how it fails \cite{boonekamp2015heuristic}, it is also a rather crude oversimplification. First, in real biological systems, the large-scale structure of interdependence network is far from ``random''. Secondly, there can be varying amounts and varying types of dependencies. In an actual complex biological system, the exchange of signals and goods between cells occurs via either diffusion or complex patterns of vascular circulation. As such,
biophysically grounded analogs of interdependence networks are necessary.

Such biophysical extensions have been investigated experimentally \cite{acun2017tissue} and theoretically \cite{suma2018interdependence} to understand how failure propagates through tissues, as mediated by the loss of diffusing cooperative factors. These cooperative factors could be cytokines (e.g. interleukin 15), growth factors (e.g. epidermal growth factor), survival factors (e.g. insulin-like growth factor 1), and antioxidant enzymes (e.g. superoxide dismutase 3) \cite{zhao2016enhanced,li2010paracrine,jiang2006supportive,lin2011cell,milan2002short,boissard2017nurse,davison2013antioxidant,fonseca2007locally,tamm1990insulin} diffusing across cells. However, the role of convective circulation, which is the primary mode of transport in large complex organisms, is missing.

The central argument of this paper is that cell failure cascades to higher structures along the circulatory network of an organism, in the same direction as the advective flow. To this end, we theoretically and experimentally study the mechanical nature of damage propagation through tissues in the presence of advective flow, and demonstrate and quantify how failures accumulate and propagate in relation to the flow direction. While our experiments are conducted on synthetic mammalian tissues in a microfluidic device, from here and with the help analytical arguments, we aim to derive more general lessons about the aging dynamics in complex organisms in vivo.

The demographic hallmark of aging is a monotonically increasing mortality rate $\mu(t)$. In strongly aging species (like humans), the probability of death of an old individual is many times larger than a young individual. Interestingly, there are phylogenetic correlations in aging characteristics \cite{jones2014diversity}: Mammalian populations have the steepest mortality curves, whereas $\mu(t)$ of amphibians and reptiles change little over time, and plants tend to age even less, and they can exhibit mortality rates that even decrease over time. The cause behind these phylogenetic trends is not entirely clear, but one possibility may be the differences in how goods and signals are transported between cells, manifesting as differences in how failures accumulate and propagate. 

For example, assaulting a sizeable portion of a shrub might not kill it, because of how little the rest of the shrub depends on the assaulted portion. The interdependence structure of a shrub is highly localized. In organisms with fast and efficient circulatory systems, however, advective transport enables any cell to depend on any other, no matter how distant. A malfunction in an animal gland or organ will affect all cells that are coupled to it via bloodstream. Thus, as much as convective flow propagates goods and signals, it should also propagate failure along the same path once the goods and signals go missing. 

In our experiments, we encapsulate rat cells in a non-degradable hydrogel (where they cannot proliferate, migrate, or contact each other), seated in an engineered microfluidic microchannel, through which we apply unidirectional flow of media, meant to emulate blood, lymph or interstitial flow. We then analyze cell death rate across the channel as a function of time and flow rate. Directing cooperative factors downstream leads to a wave of failure starting near the inlet and propagating towards the outlet, which we analytically describe using a mechanistic diffusion-convection-population-dynamics model.  

From this model we obtain relevant length scales in terms of system parameters and determine the conditions for where a failure wave should originate, and obtain the velocity and acceleration of its propagation.


\section*{Materials and Methods}
\begin{figure}
    \centering
    \includegraphics[width=0.48\textwidth]{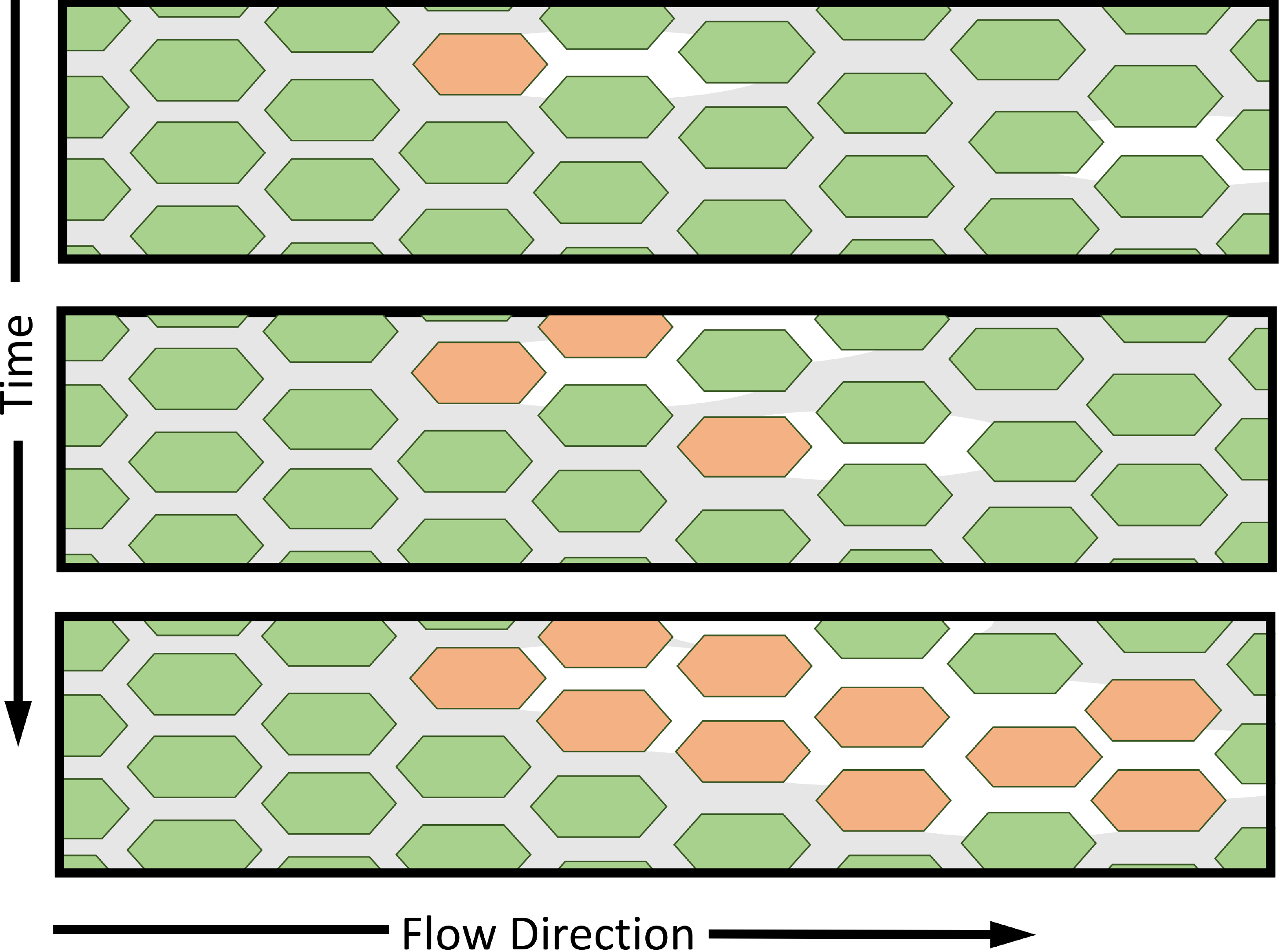}
    \caption{{\bf Schematics of proposed mechanism of tissue failure.} Healthy cells (green) secrete factors (grey) that enable or enhance the function of other cells. Advective flow transport cooperative factors downstream. Once a cell fails/dies (red), it can no longer support cells that are dependent on it. A cascade of failure then propagates over time in the direction of flow.}
    \label{fig:my_label}
\end{figure}
\subsection*{Construction of the tissue engineered model}

{\bf Fabrication of the microfluidic device.} Poly(dimethylsiloxane) (PDMS) prepolymer and curing agent (Dow Corning, USA) were mixed thoroughly at a 10:1 ratio and cast on a silicon-base mold with 20mm $\times$ 0.5mm $\times$ 0.25mm (length $\times$ width $\times$ height) ridges, and cured at 80\textcelsius{} for 30min. The PDMS and a glass slide were treated with air plasma for 1min and immediately bound to each other with the channel side facing the slide. The device was sterilized under UV for $2$h.

{\bf Preparation of the polyethylene glycol (PEG) solution.} Maleimide polyethylene glycol (PEG) succinimidyl carboxymethyl ester (PEG-NHS, Mw 3400Da, JenKem Technology) was conjugated with tyrosine-arginine-glycine-aspartic acid-serine (Y\textbf{RGD}S, Bachem) as described previously, to obtain the PEG-RGD \cite{acun2017tissue}. The PEG-RGD and 4-arm PEG-acrylate (4-PEG-ACR, 20kDa, JenKem Technology) were mixed at a 1.5:8.5 (w/w) ratio, and a 20\% (w/v) solution was prepared in culture medium (DMEM-high glucose containing 10\% fetal bovine serum (FBS) and 1\% penicillin-streptomycin). The photoinitiator Irgacure D-2959 (Sigma-Aldrich) (final concentration: 0.1\%, w/v) was added. 

\begin{figure*}
    \centering
    \includegraphics[width=0.99\textwidth]{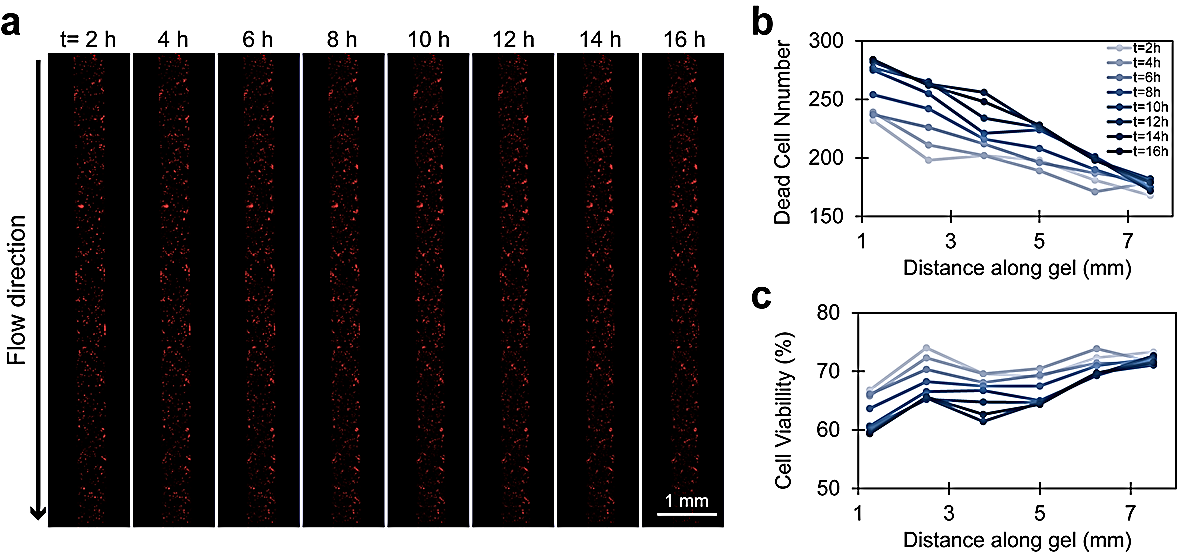}
    \caption{{\bf Time-lapse imaging of the cardiac fibroblast-laden PEG:PEG-RGD gels in the microchannel showing cell viability in real-time.} {\bf (a)} Time-lapse fluorescence microscopy images showing the dead cells in the gel every 2 h for 16h. Red: ethidium homodimer-1. {\bf (b, c)} Line graphs showing the change in cell viability along the microchannel over time. {\bf (b)} Dead cell number, and {\bf (c)} cell viability. Flow rate: 20 $\mu$L/min.}
    \label{fig:experiment}
\end{figure*}
{\bf Cell culture and seeding.}
Cardiac fibroblasts were isolated from hearts of 2-day-old Sprague-Dawley rats (Charles River Laboratories), according to the IACUC guidelines with the approval of the University of Notre Dame, which has an approved Assurance of Compliance on file with the National Institutes of Health, Office of Laboratory Animal Welfare. Rats were sacrificed via $\text{CO}_2$ treatment, and the hearts were immediately collected, minced and incubated in trypsin (Life Technologies) at 4 \textcelsius{} for 16h with gentle agitation as described previously \cite{acun2017tissue}. After further digestion with collagenase type II (Worthington-Biochem) at 37\textcelsius{}, the tissues were strained through 40$\mu$m filters and the cells in filtrate were incubated at 37\textcelsius{} for 2h. This brief incubation allowed exclusively fibroblasts to attach on the plate. After removal of unattached cells, Fibroblasts were incubated in culture media with media change every 3 days until passage 4 (P4). When the cells were at approximately 80\% confluency, they were detached from flasks using 0.25\% trypsin-EDTA (Life Technologies), counted, and reconstituted in culture medium. 

The cell suspension was mixed with the PEG:PEG-RGD solution at 1:1 volume ratio, and loaded into the microfluidic device such that the final cell density was $5\times10^6$ cells/mL, PEG:PEG-RGD concentration was 10\% (w/v), and photoinitiator concentration was 0.05\% (w/v). The polymer was exposed to UV at 365nm wavelength and 6.9mW/$\text{cm}^2$ intensity for 60s to crosslink it. The device was connected to a pump loaded with a syringe full of culture medium, and perfused at a flow rate of $150\mu$L/h for 30min to wash away the photoinitiator and the uncrosslinked polymer remaining in the gel. The bright field images of the gel were taken using a microscope (Zeiss, Hamamatsu ORCA flash 4.0), and the initial total cell number was determined using Fiji software (NIH). 

\subsection*{Cell viability measurements}

Cell viability was determined using the Live/Dead cell viability assay (Thermo Fisher Scientific) either daily (at Days 0, 1, and 2) or in real-time (every 2h for 16h). For the daily analysis, the microfluidic device was disconnected at each time point, perfused with PBS containing ethidium homodimer-1 (EthD-1) (4$\mu$M) for 30min, and imaged under bright field (total cells) or fluorescence (dead cells, red) modes with a fluorescence microscope (Zeiss, Hamamatsu ORCA flash 4.0). The device was reconnected to the syringe pump full of culture medium, and perfused at $20\mu$L/h flow rate until the next time points (Day 1 and Day 2). 

To monitor cell viability in real-time, the device was perfused with culture medium containing EthD-1 (4$\mu$M) at two flow rates (20 and $120\mu$L/h), and imaged under bright field at time 0 (total cells), and then under fluorescence with time-lapse imaging at every 2h for 16h (dead cells). The cell numbers in gel areas of 0.5mm $\times$ 1.25mm were determined all along the gel using Fiji software (NIH). Three different measurements of the same area, the second and third being at 0.2mm distances from the first one, were done to calculate the average cell numbers. The data from the middle portion of the gel were analyzed to eliminate the boundary effect. Cell viability was calculated as (Total - Dead)/\text{Total}.


\section*{Experimental results}



To investigate how failure propagates through a tissue under flow conditions, a microfluidic channel filled with cardiac fibroblast-laden hydrogel was perfused at various flow rates and imaged in real-time every 2h for 16h. At 20$\mu$L/min flow rate, we observed a gradual increase in the dead cell number over time in the inlet, while dead cell number remained relatively stable over time at the outlet (Fig. \ref{fig:experiment}a,b). The difference between the dead cell number at $t = $2h and at $t = $16h gradually decreased towards the outlet. Thus, after 16h of perfusion, cell viability was significantly higher at the outlet portion relative to the inlet portion (Fig. \ref{fig:experiment}c). 

To see the effect over a long time, in another experiment, the engineered tissue was perfused at 20$\mu$L/h flow rate for 2 days, and the dead cells were imaged and counted at Days 0, 1 and 2 (Supplementary Fig. \ref{fig:suppl_exp_one}). Upon perfusion, more cells died at the inlet and cell viability showed an increasing trend along the gel towards the outlet. Viability at the inlet was significantly lower than that at the outlet both for Day 1 ($p<0.0004$) and for Day 2 ($p<0.0008$).



Considering that the flow rate did not change along the microchannel, the shear stress exerted on the cells by the flow itself should be the same. Thus, we concluded that the gradual increase in cell viability towards the outlet was not because of the shear force, but rather due to accumulation of the cooperative factors. 

\begin{figure}
    \centering
    \includegraphics[width=0.49\textwidth]{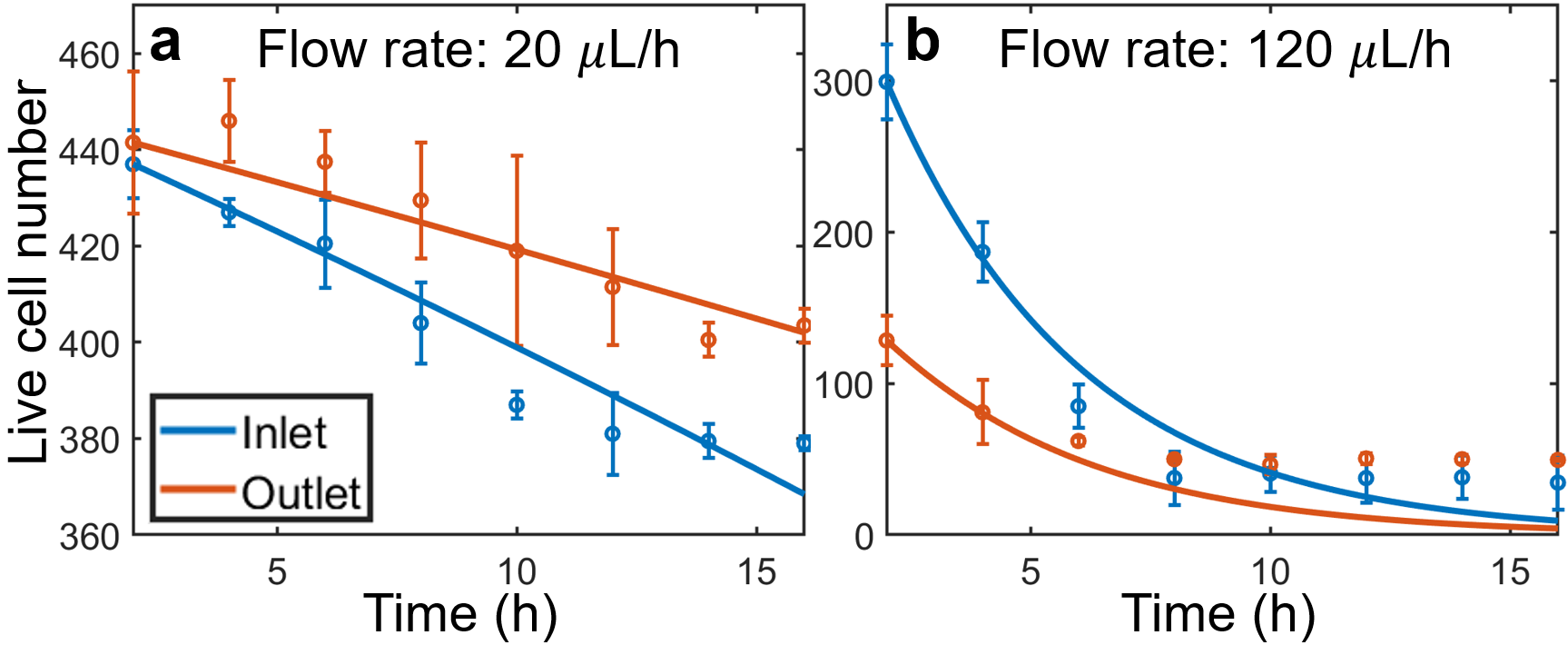}
    \caption{{\bf Experimental data with fitted curves.}
    Fitted parameter values are $k = 1.87$, $\alpha = 0.25$, $\beta_{\text{inlet}} = 5.30$, $\beta_{\text{outlet}} = 7.13$ for $v=$ 20 $\mu$L/h, and $\beta_{\text{inlet}} = 1.43 \times 10^{-14}$, $\beta_{\text{outlet}} = 0.27$ for $v=$ 120 $\mu$L/h. We indeed see $\beta_{\text{inlet}}$ < $\beta_{\text{outlet}}$ for each flow rate, signifying a larger portion of cooperative factors in the vicinity of cells at the outlet. Inlet cell populations were given by averaging over measurements taken between 1.25 and 2.5 mm from the inlet for $v=$ 20 $\mu$L/h and 3.75 and 5 mm from the inlet for $v=$ 120 $\mu$L/h. Outlet cell populations were given by averaging over measurements taken between 5 and 6.25 mm from the inlet for $v=$ 20 $\mu$L/h and 8.75 and 10 mm from the inlet  for $v=$ 120 $\mu$L/h. Error bars correspond to one standard deviation from the mean. }
    \label{fig:exp_fit}
\end{figure}

The difference between the relative cell viability (viability at a given time relative to that at t=2h) at the inlet and outlet changed dramatically when the flow rate was altered (Fig. \ref{fig:exp_fit}). At 20$\mu$L/min flow rate, relative cell viability decreased by 2\% at the outlet and 14\% at the inlet over a 16h period, with a gradual decrease at both edges (Fig. \ref{fig:exp_fit}a). The viability difference between the inlet and outlet reached 12\% at $t=$16h. At 120$\mu$L/min flow rate, viability at the outlet decreased by 58\% over a 16h period, while that at the inlet decreased by 93\% (Fig. \ref{fig:exp_fit}b). Although cell viability was initially higher at the inlet (69\%) than the outlet (38\%), it decreased dramatically over time and became higher at the outlet (16\%) than the inlet (11\%). 

Therefore, at the low flow rate, cell viability was the highest, but the difference between the viability at the inlet and outlet was the smallest. With the increasing flow rate, viability decreased remarkably especially at the inlet but also at the outlet. These observatıons suggest that, at lower flow rates, cooperative factors were not completely cleared from the inlet and could bind to their receptors on cell surfaces, improving cell viability outcomes. When the flow rate was increased, more cooperative factors were washed out from the inlet, giving them less time to bind to receptors on the cells and dramatically decreasing cell viability. As cells at the inlet die out, they are no longer able to produce cooperative factors to aid the survival of cells at the outlet, leading to a propagation of failure and a reduced viability of cells at the outlet.

\section*{Analytical results}
\subsection*{Model}
To characterize the dynamics of flow mediated failure propagation in a more general setting, and to make predictions beyond the experiments described above, we develop an analytical model the inputs of which we obtain from experiments described above. Our model assumptions, qualitatively stated, are as follows. (1) The cells do not proliferate and migrate, but are under stress and die. We denote the cell density at position $x$ along the channel at time $t$, with $n(x,t)$. (2) Cells secrete some cooperative factor(s) that will diffuse, flow and decay within the circulating fluid, and help other cells survive/function. We assume that the diffusion and decay parameters for these factors are similar, and denote their concentration collectively by a single quantity $\Phi(x,t)$. These assumptions can be quantified as
\begin{align}
    &\dot{n} = -\alpha \frac{\phi_0^k}{\phi_0^k + \Phi^k} n \label{eq:cells} \\
    &\dot{\Phi} = d \nabla^2 \Phi - \mathbf{v}\cdot \boldsymbol{\nabla} \Phi - \gamma \Phi + A n .
    \label{eq:chemicals}     
\end{align}
where, $d$ is the diffusion constant for the cooperative factors, $\mathbf{v}$ is the flow velocity, $A$ is the rate at which cells secrete cooperative factors, $k$ describes the steepness of the response of cells to cooperative factors, and the constant $\phi_0$ quantifies the ``required amount'' of factors for a cell to function normally. The functional form of the right hand side of eqn.\ref{eq:cells} is the Hill function, which accurately describes a cell's response to many different kinds of molecular agents. It is motivated by Michaelis-Menten type reaction kinetics, fits experimental findings, and is ubiquitously used in population dynamics models. $\alpha$ is the cell death rate when there are no cooperative factors. In a variable environment $\alpha$ could be time dependent, however here we take it to be constant.

We study the phenomena of failure propagation by first extracting the relevant length scales. From eqn.(\ref{eq:chemicals}), we can solve the Green's function for a point source (Appendix \ref{sec:greens}). From the Green's function, we can then extract left $\ell_{L}$ and right $\ell_{R}$ length scales given as, 
\begin{equation}
    \label{eq:lengths}
    \ell_{L} (v) = \left[ \lambda(v) + \frac{v}{2d} \right]^{-1}, \quad \ell_{R} (v) = \left[ \lambda(v) - \frac{v}{2d}  \right]^{-1},
\end{equation}
where we have defined $\lambda(v) = \sqrt{ \gamma / d + v^2 / 4 d^2 }$ and flow is taken to be from left to right $(v \geq 0)$. These length scales correspond to the characteristic advection-diffusion-decay length of the cooperative factors given by a point source. 

As the flow velocity $v$ increases, the left length goes to zero and the right length increases roughly linearly with respect to $v$. This introduces a bias in the concentration of cooperative factors, and we expect to see a propagation of cell death in the direction of flow. As cells upstream die, the cooperative factors upstream also diminish, causing a propagation of death moving downstream, in the direction of flow. 


\begin{figure}
    \centering
    \includegraphics[width=0.48\textwidth]{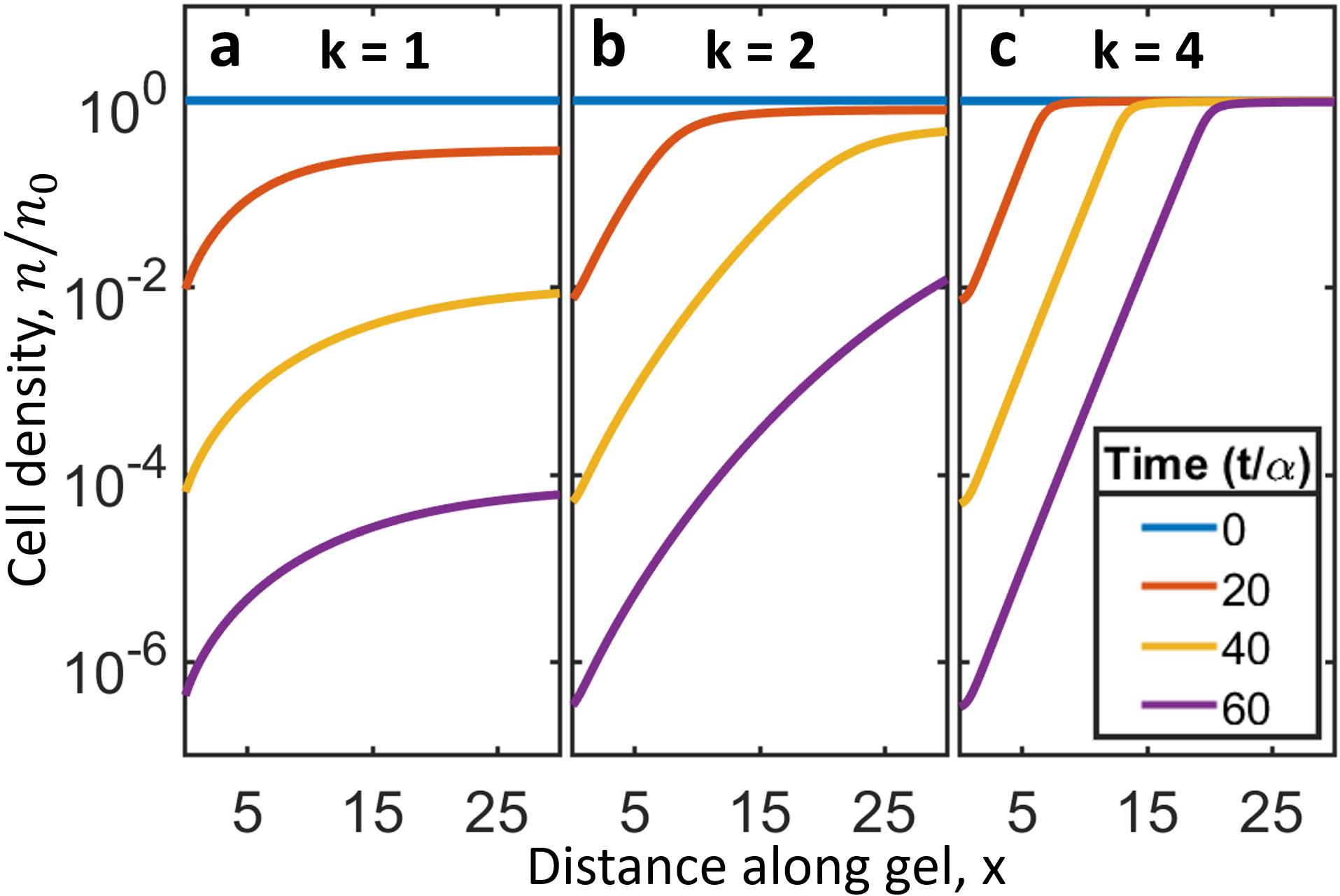}
    \caption{{\bf Simulation results for various regimes.} If the initial cell concentration is sufficiently large ($n_0 > \gamma \phi_0 /A$), there will be a propagation of death due to flow. Here we numerically solve and plot cell concentration normalized by the initial concentration at various times normalized by the death rate $\alpha$. {\bf (a)} For low hill constant ($k=1$), we see a permanent gradient formed before the bulk goes below the critical value, $n_0 < \gamma \phi_0 /A$, and cells begin to die uniformly. {\bf (b)} As we increase the hill constant ($k=2$), the bulk cell density attenuates slower and a more pronounced wave of failure begins to develop. {\bf (c)} In the case of a strong response function ($k=4$), we see a more pronounced wave of propagating failure. Here, the bulk concentration remains roughly constant and death propagates from the inlet to outlet, along direction of flow, at roughly a constant death velocity. }
    \label{fig:sim_wave}
\end{figure}
\subsection*{Experimental fits}
To fit our model to experimental data, we first make some simplifying assumptions, in order to reduce our model to a solvable system. We assume the local cooperative factor concentration $\Phi(x,t)$ will be proportional to the local cell density $n(x,t)$. We denote this proportionality constant $\beta$, so that $\Phi(x,t) = \beta n(x,t)$. In the center of the microchannel, where $n$ is roughly uniform, this constant would be given by taking the steady state cooperative factor concentration, giving $\Phi = A n / \gamma$, so $\beta = A /\gamma$. By the inlet, this constant will be lower in value since the area of cells contributing to the cooperative factors will be reduced. 

The constant $\beta$ will in general depend on the length scales contributing to the local cooperative factor concentration and thus will vary between the inlet and outlet regions. Specifically, we expect the value of $\beta_{\mathrm{inlet}}$ to scale as $ A \ell_L /(\ell_L + \ell_R) \gamma$ and $\beta_{\mathrm{outlet}} \sim A \ell_R /(\ell_L + \ell_R) \gamma$. Therefore, for a large flow rate, we expect the value of $\beta_{\mathrm{inlet}}$ to be less than $\beta_{\mathrm{outlet}}$ since more cells are able to contribute to the cooperative factor concentration at the outlet. With this assumption, we can write down an effective equation describing the growth of the cells over time as,
\[
\dot{n} = \frac{-\alpha n}{1+(\beta n/n_0)^k},
\]
where $n_0$ is the initial cell concentration. We can then solve this exactly to get,
\[
 n(t) = (n_0/\beta) \left[ W\left(e^{\beta^k -\alpha k t} \beta^k \right) \right]^{1/k} 
\]
where $W(z)$ is the Lambert W function, defined as the principle solution for $w$ in $z = w e^w$ and can be computed to arbitrary numerical precision.

We then fit our model to experimental data in Fig.\ref{fig:exp_fit}. Since we expect the growth kinetics to be the same in both experiments, and the only difference being flow, we constrain the fit parameters for $\alpha$ and $k$ to be the same for all fits and allow different values of $\beta$ for inlet and outlet regions and for each flow rate $v=$ 20$\mu$L/h and $v=120\mu$L/h. For hill and decay constants we then get $k = 1.87$, $\alpha = 0.25\mathrm{hr}^{-1}$. For flow rate $v=$ 20$\mu$L/h, we get $\beta_{\text{inlet}}=5.30$, $\beta_{\text{outlet}} = 7.13$. For $v=120\mu$L/h, we get $\beta_{\text{inlet}} = 1.43 \times 10^{-14}$, $\beta_{\text{outlet}} = 0.27$. We see that indeed $\beta_{\mathrm{inlet}} < \beta_{\mathrm{outlet}}$ as expected for each flow rate. We also see the values of $\beta$ are much smaller for the larger flow rate, with the inlet value of $\beta$ being vanishingly small at the flow rate $v=120\mu$L/h. This is because with with large flow rate, the cooperative factors are pushed much further downstream and no longer help the cells at the inlet and may also diminish the cooperative factor concentration at the outlet.

We next study the system of equations (\ref{eq:cells}-\ref{eq:chemicals}) and derive conditions for a propagation of failure, as well as the velocity and acceleration of failure propagation.

\subsection*{Model regimes and failure propagation}

From numerical simulations and dimensional analysis, we find the condition for failure propagation to occur is that the initial density of cells $n_0$ must be sufficiently large, such that $A n_0/\gamma \gg \phi_0$. This is since the largest the cooperative factor concentration can be is given by $A n_0/\gamma$ and this value must be above the threshold concentration $\phi_0$ for cell density to not decay exponentially. If the cell concentration drops below this critical value, there will no longer be a pronounced propagation of failure. Instead, the entire population of cells will die roughly uniformly at an exponential rate of $\alpha$. 

For cases where there is a propagation of failure, where $A n_0 / \gamma > \phi_0$, we plot the numerical solution to our system in Fig. \ref{fig:sim_wave}, for values of the hill constant $k=1,2,4$. We find that as we increase the value of the hill constant, we get a more pronounced wave. The population dynamics is strongly determined by the initial density $n_0$ and the form of the response given by the hill constant $k$. In the case where $k$ is low, the bulk where $\Phi(x) > \phi_0$ will attenuate quicker. We find this attenuation of the bulk leads to an ``acceleration'' of death. We also study this attenuation and the corresponding acceleration of death propagation.

We now determine the velocity at which the death of cells will propagate. Our derivations for failure propagation velocity, depth, and acceleration are given in detail in Appendix \ref{sec:derivations}. We illustrate here the procedure used in our derivations. 

To simplify our analysis, we used a boxcar approximation to the Green's function for the cooperative factors where left and right lengths are given by equations (\ref{eq:lengths}), and assume chemicals quickly reach steady state. The total area of the Green's function corresponds to the secretion rate per cell density at steady state. We therefore have $G(x) = (A/\gamma) \Theta(x + \ell_{L} ) \Theta( \ell_{R} -x ) $, where $\Theta(x)$ is the Heaviside step function. 

We then convolve this boxcar Green's function with a semi-infinite initial cell concentration, $n(x,0) = n_0 \Theta(x)$. This gives a cooperative factor concentration profile that increases linearly up until $x = \ell_R$, after which $\Phi(x)$ saturates to a constant given by $\Phi(x > \ell_R) = A n_0 /\gamma$. Assuming the initial cell density is sufficiently large such that $A n_0 / \gamma > \phi_0$, we can find the position $\Delta$, such that $\Phi (\Delta) = \phi_0$. To the left of this value, cells are expected to decay exponentially at a rate $\alpha$. The cooperative factor concentration will then update to this new concentration of cells and we can reiterate this to get the next decay of cells. The death of cells will therefore continue propagate by an amount $\Delta$ at a rate $\alpha$, giving a first approximation to the failure propagation velocity as $v_d = \alpha \Delta$. We find this gives good agreement for large $k$, where the cell death rate behaves closer to a step function, and for short times. 

We then further improve these calculations by performing a second iteration with an updated approximate cell concentration, taking into account the region where $\phi_0 < \Phi(x) < A n_0 / \gamma$, as well as taking into account the attenuation of the bulk $n_b(t) = n(x > \ell_R, t)$. Details of this procedure are given the Appendix \ref{sec:derivations}. Our final result for the velocity of failure propagation is,
\begin{equation}
\label{eq:death_velocity}
 v_d = \frac{\alpha v}{2 \gamma} + \frac{\alpha \sqrt{v^2 + 4 d \gamma}}{2 \gamma (1 + 2^k)} \left[ 1 - 2^k \!+\!  2^{1+k} \left( \frac{u}{1 - \alpha k t u} \right)^{\frac{1}{k}}\right]\nonumber
\end{equation}
where $u=[\gamma \phi_0/(An_0)]^k$. The initial cell density will then see a failure propagation at the inlet end at this velocity, as well as an attenuation of the bulk density $n_b(t)$. For the evolution of the bulk density, we note that $\Phi(x)$ in this region is given as $A n_b(t) / \gamma$. We then expand the Hill function in equation (\ref{eq:cells}) about infinity (for $A n_b(t) / \gamma \gg \phi_0$) and solve for $n_b(t)$. We then can get for the bulk density,
\begin{equation}
    \label{eq:bulk}
    n_b (t) = n_0(1 - \alpha u k t)^{1/k}.
\end{equation}

\begin{figure}
    \centering
    \includegraphics[width=0.48\textwidth]{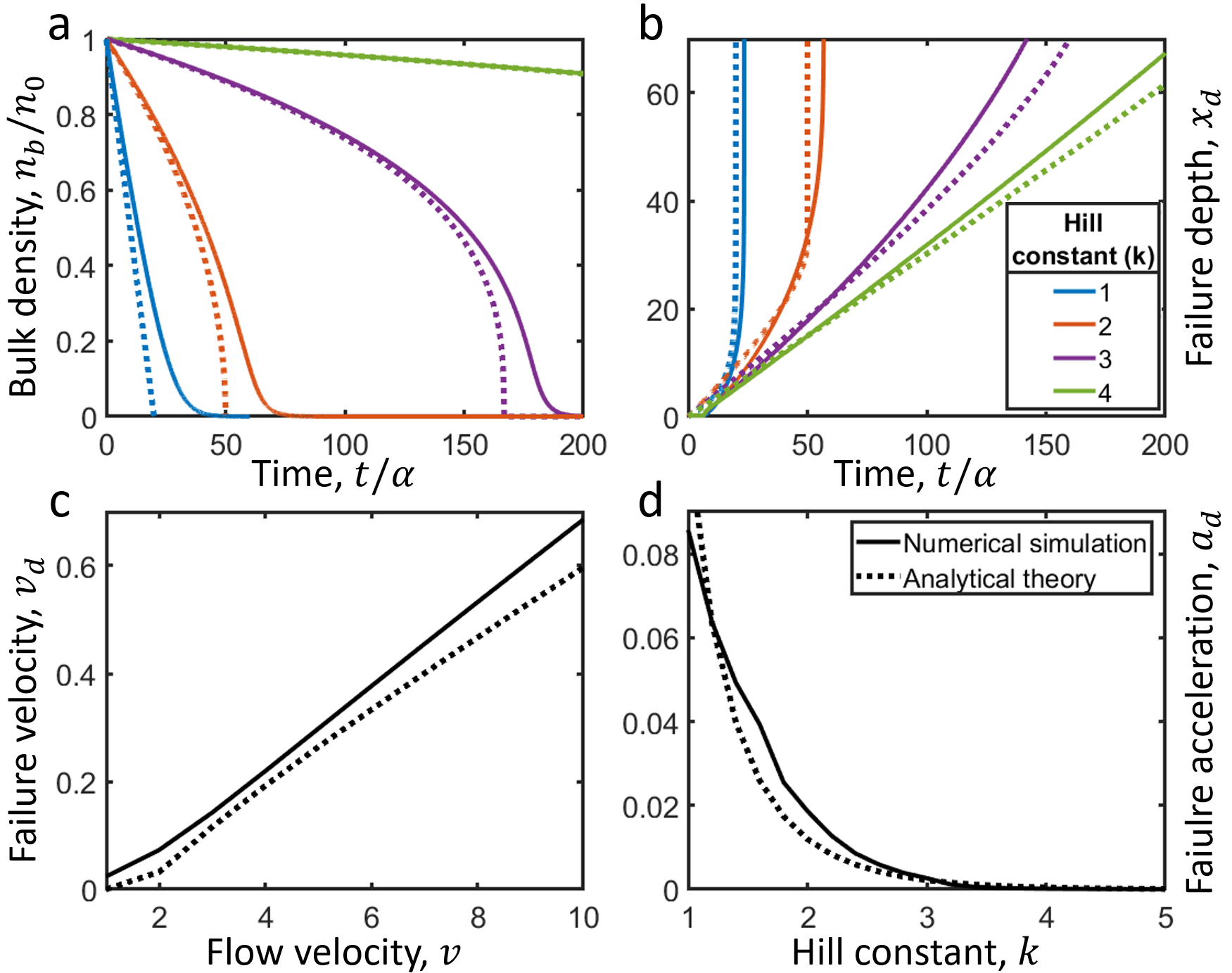}
    \caption{{\bf Analytical theory and numerical simulations for failure propagation and bulk death versus time.} {\bf (a)} Bulk cell concentration normalized by initial cell concentration $n_b/n_0$, versus time. For lower values of the hill constant $k$, the cell concentration in the bulk decays quicker. At a critical time $t_c$, the bulk density falls below the critical value $n_b < \gamma \phi_0 /A$ and dies out exponentially. {\bf (b)} Propagation of failure versus time for hill constants $k=1,2,3,4$. For lower hill constants, the failure depth rapidly increases up to a critical time $t_c$, where the bulk collapses before the propagating wave reaches the end of the domain. For larger hill constants, the velocity remains roughly constant and reaches the end of the bulk before time $t_c$. {\bf (c)} Failure propagation velocity vs flow velocity. The failure propagation velocity $v_d$ increases roughly linearly with the flow velocity $v$. {\bf (d)} Failure propagation acceleration vs hill constant. As the hill constant $k$ increases, the failure propagation becomes more pronounced and moves at a more constant speed. The acceleration then decreases roughly exponentially as k increases. Solid lines are obtained from numerical simulations throughout and dotted lines are obtained from analytical theory. }
    \label{fig:model_dynamics}
\end{figure}

We compare our results with numerical simulations for the failure penetration depth and bulk attentuation over time for values of $k = 1$ to $4$ in Fig. \ref{fig:model_dynamics}a,b and see good agreement with analytical formulas.

Note that the velocity diverges when the term in the inner most square bracket vanishes in equation (\ref{eq:death_velocity}). This happens at a critical time,
\begin{equation}
    t_c = 1/(\alpha k u).
\end{equation}
This corresponds to the time at which the bulk goes below the initial critical cell density, that is, when $n_b(t)$ from equation (\ref{eq:bulk}) vanishes. After this there is no more pronounced wave. From the expression for the death propagation velocity, we can also determine a death depth. Depending on the tissue length $L$, this death depth may occur before or after the time the bulk collapses. 

The attenuation of the bulk also leads to an ``acceleration'' of failure propagation. This effect is largest for low $k$, since the bulk decays faster as the Hill function saturates slower. We can get this acceleration, $a_{d}$ by taking a time derivative of $v_{d}$,
\begin{equation}
    a_d = \frac{ \alpha^2 \sqrt{v^2 + 4 d \gamma}}{\gamma (2^{-k} + 1) } \left(\frac{u}{ 1- \alpha u k t} \right)^{ \frac{k + 1}{k} } . \label{eq:death_acceleration}
\end{equation}
Since $A n_0 / \gamma \gg \phi_0$, we see as $k \rightarrow \infty$, the first term in the square brackets grow much faster than the linear term $\alpha k t$. Since the exponent is overall negative, this corresponds to a large positive term in the denominator as $k \rightarrow \infty$, and the death acceleration goes to zero. 

We compare our analytical results for the velocity and acceleration of death with numerical simulations and get good agreement (Fig \ref{fig:model_dynamics}c,d).


\section*{Discussion}
Here we studied how failure propagates in a system where interdependence is mediated by flow. These results emphasize the importance of intercellular processes on aging.

We performed experiments with synthetic tissues filled with cardiac fibriblast-laden PEG hydrogels in a microchannel and observed that flow can help increase the lifespan of cells downstream of the flow (Fig.1, 2). We explained this observation with cooperative factors, which were carried by the flow towards the outlet. Cooperative factors are known to promote cell survival \cite{suma2018interdependence,acun2017tissue,zhao2016enhanced,li2010paracrine,jiang2006supportive,lin2011cell,milan2002short,boissard2017nurse}, cardioprotection \cite{zhao2016enhanced,li2010paracrine}, and angiogenesis \cite{li2010paracrine,jiang2006supportive}, and cells failing to receive the necessary factors from the neighboring cells go through apoptosis \cite{petralia2014aging}. Motivated by these results, we developed an analytical model to describe the death of cells that communicate via diffusive cooperative factors in a flowing environment.

Fitting this model to experiment, we saw indeed that the proportion of cooperative factors downstream of the flow were larger than those upstream. This then leads to a faster death of cells by the inlet and longer lifespan for cells by the outlet.

Next, we investigated further the consequences of our analytical model. We found analytical conditions for a ``wave'' of failure propagation to occur in the direction of flow. As cells die out upstream, this failure will propagate and increase the mortality rate of cells downstream. The conditions for this propagation to occur were found to be a sufficient density of cells and a non-zero flow rate. Once cell density decreases below a critical threshold, the cells will die uniformly at roughly the same exponential rate.

In the case of sufficiently large cell density, we see the form of failure propagation is heavily dependent on the hill constant $k$. For smaller values of $k$, cells that are further downstream die faster compared to large $k$, leading to an ``acceleration'' of failure. For large constants $k$, the wave front is more pronounced and continues at more of a constant velocity.

Through a dimensional analysis and simplifying assumptions, we were able to derive analytical formulas describing the propagation depth, velocity, and acceleration of failure, as well as the attenuation of cells beyond the wave front.

Failure propagation has already been studied previously when cells are coupled through diffusion \cite{suma2018interdependence,acun2017tissue}. Here, we analyzed the failure propagation under flow, and developed an analytical model that describes the depth, velocity, and acceleration of failure propagation. This is physiologically more relevant, since cooperative factors are carried via flow in the body. For example, paracrine factors are transported via interstitial flow \cite{yao2012interstitial,swartz2012lymphatic}. 

Fluid flow is known to mediate communications within microbial communities and influence death patterns \cite{rossy2019cellular}. Thus, while our model was motivated by experiments on mammalian tissues, we might expect similar results to also hold for eukaryotic colonies and bacterial biofilms where diffusive and advective forces are responsible for the communication of interdependent members of the population.  





\section*{Author Contributions}

GU, GB, PZ and DCV formulated the problem. GB and PZ designed and carried out the experiments. GU and DCV designed and carried out the theory. GU, GB, DCV wrote the paper.

\section*{Acknowledgments}

This study was funded by National Science Foundation grant CBET-1805157


\bibliography{bibliography}

\begin{thebibliography}{28}%
\makeatletter
\providecommand \@ifxundefined [1]{%
 \@ifx{#1\undefined}
}%
\providecommand \@ifnum [1]{%
 \ifnum #1\expandafter \@firstoftwo
 \else \expandafter \@secondoftwo
 \fi
}%
\providecommand \@ifx [1]{%
 \ifx #1\expandafter \@firstoftwo
 \else \expandafter \@secondoftwo
 \fi
}%
\providecommand \natexlab [1]{#1}%
\providecommand \enquote  [1]{``#1''}%
\providecommand \bibnamefont  [1]{#1}%
\providecommand \bibfnamefont [1]{#1}%
\providecommand \citenamefont [1]{#1}%
\providecommand \href@noop [0]{\@secondoftwo}%
\providecommand \href [0]{\begingroup \@sanitize@url \@href}%
\providecommand \@href[1]{\@@startlink{#1}\@@href}%
\providecommand \@@href[1]{\endgroup#1\@@endlink}%
\providecommand \@sanitize@url [0]{\catcode `\\12\catcode `\$12\catcode
  `\&12\catcode `\#12\catcode `\^12\catcode `\_12\catcode `\%12\relax}%
\providecommand \@@startlink[1]{}%
\providecommand \@@endlink[0]{}%
\providecommand \url  [0]{\begingroup\@sanitize@url \@url }%
\providecommand \@url [1]{\endgroup\@href {#1}{\urlprefix }}%
\providecommand \urlprefix  [0]{URL }%
\providecommand \Eprint [0]{\href }%
\providecommand \doibase [0]{http://dx.doi.org/}%
\providecommand \selectlanguage [0]{\@gobble}%
\providecommand \bibinfo  [0]{\@secondoftwo}%
\providecommand \bibfield  [0]{\@secondoftwo}%
\providecommand \translation [1]{[#1]}%
\providecommand \BibitemOpen [0]{}%
\providecommand \bibitemStop [0]{}%
\providecommand \bibitemNoStop [0]{.\EOS\space}%
\providecommand \EOS [0]{\spacefactor3000\relax}%
\providecommand \BibitemShut  [1]{\csname bibitem#1\endcsname}%
\let\auto@bib@innerbib\@empty
\bibitem [{\citenamefont {Rattan}(2006)}]{rattan2006theories}%
  \BibitemOpen
  \bibfield  {author} {\bibinfo {author} {\bibfnamefont {S.~I.}\ \bibnamefont
  {Rattan}},\ }\href@noop {} {\bibfield  {journal} {\bibinfo  {journal} {Free
  radical research}\ }\textbf {\bibinfo {volume} {40}},\ \bibinfo {pages}
  {1230} (\bibinfo {year} {2006})}\BibitemShut {NoStop}%
\bibitem [{\citenamefont {Blackburn}(2000)}]{blackburn2000telomere}%
  \BibitemOpen
  \bibfield  {author} {\bibinfo {author} {\bibfnamefont {E.~H.}\ \bibnamefont
  {Blackburn}},\ }\href@noop {} {\bibfield  {journal} {\bibinfo  {journal}
  {Nature}\ }\textbf {\bibinfo {volume} {408}},\ \bibinfo {pages} {53}
  (\bibinfo {year} {2000})}\BibitemShut {NoStop}%
\bibitem [{\citenamefont {Murabito}\ \emph {et~al.}(2012)\citenamefont
  {Murabito}, \citenamefont {Yuan},\ and\ \citenamefont
  {Lunetta}}]{murabito2012search}%
  \BibitemOpen
  \bibfield  {author} {\bibinfo {author} {\bibfnamefont {J.~M.}\ \bibnamefont
  {Murabito}}, \bibinfo {author} {\bibfnamefont {R.}~\bibnamefont {Yuan}}, \
  and\ \bibinfo {author} {\bibfnamefont {K.~L.}\ \bibnamefont {Lunetta}},\
  }\href@noop {} {\bibfield  {journal} {\bibinfo  {journal} {Journals of
  Gerontology Series A: Biomedical Sciences and Medical Sciences}\ }\textbf
  {\bibinfo {volume} {67}},\ \bibinfo {pages} {470} (\bibinfo {year}
  {2012})}\BibitemShut {NoStop}%
\bibitem [{\citenamefont {Shay}\ and\ \citenamefont
  {Wright}(2000)}]{shay2000hayflick}%
  \BibitemOpen
  \bibfield  {author} {\bibinfo {author} {\bibfnamefont {J.~W.}\ \bibnamefont
  {Shay}}\ and\ \bibinfo {author} {\bibfnamefont {W.~E.}\ \bibnamefont
  {Wright}},\ }\href@noop {} {\bibfield  {journal} {\bibinfo  {journal} {Nature
  reviews Molecular cell biology}\ }\textbf {\bibinfo {volume} {1}},\ \bibinfo
  {pages} {72} (\bibinfo {year} {2000})}\BibitemShut {NoStop}%
\bibitem [{\citenamefont {Gavrilov}\ and\ \citenamefont
  {Gavrilova}(2001)}]{gavrilov2001reliability}%
  \BibitemOpen
  \bibfield  {author} {\bibinfo {author} {\bibfnamefont {L.~A.}\ \bibnamefont
  {Gavrilov}}\ and\ \bibinfo {author} {\bibfnamefont {N.~S.}\ \bibnamefont
  {Gavrilova}},\ }\href@noop {} {\bibfield  {journal} {\bibinfo  {journal}
  {Journal of theoretical Biology}\ }\textbf {\bibinfo {volume} {213}},\
  \bibinfo {pages} {527} (\bibinfo {year} {2001})}\BibitemShut {NoStop}%
\bibitem [{\citenamefont {Vural}\ \emph {et~al.}(2014)\citenamefont {Vural},
  \citenamefont {Morrison},\ and\ \citenamefont {Mahadevan}}]{vural2014aging}%
  \BibitemOpen
  \bibfield  {author} {\bibinfo {author} {\bibfnamefont {D.~C.}\ \bibnamefont
  {Vural}}, \bibinfo {author} {\bibfnamefont {G.}~\bibnamefont {Morrison}}, \
  and\ \bibinfo {author} {\bibfnamefont {L.}~\bibnamefont {Mahadevan}},\
  }\href@noop {} {\bibfield  {journal} {\bibinfo  {journal} {Physical Review
  E}\ }\textbf {\bibinfo {volume} {89}},\ \bibinfo {pages} {022811} (\bibinfo
  {year} {2014})}\BibitemShut {NoStop}%
\bibitem [{\citenamefont {Vaupel}\ \emph {et~al.}(1998)\citenamefont {Vaupel},
  \citenamefont {Carey}, \citenamefont {Christensen}, \citenamefont {Johnson},
  \citenamefont {Yashin}, \citenamefont {Holm}, \citenamefont {Iachine},
  \citenamefont {Kannisto}, \citenamefont {Khazaeli}, \citenamefont {Liedo}
  \emph {et~al.}}]{vaupel1998biodemographic}%
  \BibitemOpen
  \bibfield  {author} {\bibinfo {author} {\bibfnamefont {J.~W.}\ \bibnamefont
  {Vaupel}}, \bibinfo {author} {\bibfnamefont {J.~R.}\ \bibnamefont {Carey}},
  \bibinfo {author} {\bibfnamefont {K.}~\bibnamefont {Christensen}}, \bibinfo
  {author} {\bibfnamefont {T.~E.}\ \bibnamefont {Johnson}}, \bibinfo {author}
  {\bibfnamefont {A.~I.}\ \bibnamefont {Yashin}}, \bibinfo {author}
  {\bibfnamefont {N.~V.}\ \bibnamefont {Holm}}, \bibinfo {author}
  {\bibfnamefont {I.~A.}\ \bibnamefont {Iachine}}, \bibinfo {author}
  {\bibfnamefont {V.}~\bibnamefont {Kannisto}}, \bibinfo {author}
  {\bibfnamefont {A.~A.}\ \bibnamefont {Khazaeli}}, \bibinfo {author}
  {\bibfnamefont {P.}~\bibnamefont {Liedo}},  \emph {et~al.},\ }\href@noop {}
  {\bibfield  {journal} {\bibinfo  {journal} {Science}\ }\textbf {\bibinfo
  {volume} {280}},\ \bibinfo {pages} {855} (\bibinfo {year}
  {1998})}\BibitemShut {NoStop}%
\bibitem [{\citenamefont {Caughley}(1966)}]{caughley1966mortality}%
  \BibitemOpen
  \bibfield  {author} {\bibinfo {author} {\bibfnamefont {G.}~\bibnamefont
  {Caughley}},\ }\href@noop {} {\bibfield  {journal} {\bibinfo  {journal}
  {Ecology}\ }\textbf {\bibinfo {volume} {47}},\ \bibinfo {pages} {906}
  (\bibinfo {year} {1966})}\BibitemShut {NoStop}%
\bibitem [{\citenamefont {Stroustrup}\ \emph {et~al.}(2016)\citenamefont
  {Stroustrup}, \citenamefont {Anthony}, \citenamefont {Nash}, \citenamefont
  {Gowda}, \citenamefont {Gomez}, \citenamefont {L{\'o}pez-Moyado},
  \citenamefont {Apfeld},\ and\ \citenamefont
  {Fontana}}]{stroustrup2016temporal}%
  \BibitemOpen
  \bibfield  {author} {\bibinfo {author} {\bibfnamefont {N.}~\bibnamefont
  {Stroustrup}}, \bibinfo {author} {\bibfnamefont {W.~E.}\ \bibnamefont
  {Anthony}}, \bibinfo {author} {\bibfnamefont {Z.~M.}\ \bibnamefont {Nash}},
  \bibinfo {author} {\bibfnamefont {V.}~\bibnamefont {Gowda}}, \bibinfo
  {author} {\bibfnamefont {A.}~\bibnamefont {Gomez}}, \bibinfo {author}
  {\bibfnamefont {I.~F.}\ \bibnamefont {L{\'o}pez-Moyado}}, \bibinfo {author}
  {\bibfnamefont {J.}~\bibnamefont {Apfeld}}, \ and\ \bibinfo {author}
  {\bibfnamefont {W.}~\bibnamefont {Fontana}},\ }\href@noop {} {\bibfield
  {journal} {\bibinfo  {journal} {Nature}\ }\textbf {\bibinfo {volume} {530}},\
  \bibinfo {pages} {103} (\bibinfo {year} {2016})}\BibitemShut {NoStop}%
\bibitem [{\citenamefont {Farrell}\ \emph {et~al.}(2016)\citenamefont
  {Farrell}, \citenamefont {Mitnitski}, \citenamefont {Rockwood},\ and\
  \citenamefont {Rutenberg}}]{farrell2016network}%
  \BibitemOpen
  \bibfield  {author} {\bibinfo {author} {\bibfnamefont {S.~G.}\ \bibnamefont
  {Farrell}}, \bibinfo {author} {\bibfnamefont {A.~B.}\ \bibnamefont
  {Mitnitski}}, \bibinfo {author} {\bibfnamefont {K.}~\bibnamefont {Rockwood}},
  \ and\ \bibinfo {author} {\bibfnamefont {A.~D.}\ \bibnamefont {Rutenberg}},\
  }\href@noop {} {\bibfield  {journal} {\bibinfo  {journal} {Physical Review
  E}\ }\textbf {\bibinfo {volume} {94}},\ \bibinfo {pages} {052409} (\bibinfo
  {year} {2016})}\BibitemShut {NoStop}%
\bibitem [{\citenamefont {Mitnitski}\ \emph {et~al.}(2017)\citenamefont
  {Mitnitski}, \citenamefont {Rutenberg}, \citenamefont {Farrell},\ and\
  \citenamefont {Rockwood}}]{mitnitski2017aging}%
  \BibitemOpen
  \bibfield  {author} {\bibinfo {author} {\bibfnamefont {A.}~\bibnamefont
  {Mitnitski}}, \bibinfo {author} {\bibfnamefont {A.}~\bibnamefont
  {Rutenberg}}, \bibinfo {author} {\bibfnamefont {S.}~\bibnamefont {Farrell}},
  \ and\ \bibinfo {author} {\bibfnamefont {K.}~\bibnamefont {Rockwood}},\
  }\href@noop {} {\bibfield  {journal} {\bibinfo  {journal} {Biogerontology}\
  }\textbf {\bibinfo {volume} {18}},\ \bibinfo {pages} {433} (\bibinfo {year}
  {2017})}\BibitemShut {NoStop}%
\bibitem [{\citenamefont {Boonekamp}\ \emph {et~al.}(2015)\citenamefont
  {Boonekamp}, \citenamefont {Briga},\ and\ \citenamefont
  {Verhulst}}]{boonekamp2015heuristic}%
  \BibitemOpen
  \bibfield  {author} {\bibinfo {author} {\bibfnamefont {J.~J.}\ \bibnamefont
  {Boonekamp}}, \bibinfo {author} {\bibfnamefont {M.}~\bibnamefont {Briga}}, \
  and\ \bibinfo {author} {\bibfnamefont {S.}~\bibnamefont {Verhulst}},\
  }\href@noop {} {\bibfield  {journal} {\bibinfo  {journal} {Experimental
  gerontology}\ }\textbf {\bibinfo {volume} {71}},\ \bibinfo {pages} {95}
  (\bibinfo {year} {2015})}\BibitemShut {NoStop}%
\bibitem [{\citenamefont {Acun}\ \emph {et~al.}(2017)\citenamefont {Acun},
  \citenamefont {Vural},\ and\ \citenamefont {Zorlutuna}}]{acun2017tissue}%
  \BibitemOpen
  \bibfield  {author} {\bibinfo {author} {\bibfnamefont {A.}~\bibnamefont
  {Acun}}, \bibinfo {author} {\bibfnamefont {D.~C.}\ \bibnamefont {Vural}}, \
  and\ \bibinfo {author} {\bibfnamefont {P.}~\bibnamefont {Zorlutuna}},\
  }\href@noop {} {\bibfield  {journal} {\bibinfo  {journal} {Scientific
  reports}\ }\textbf {\bibinfo {volume} {7}},\ \bibinfo {pages} {1} (\bibinfo
  {year} {2017})}\BibitemShut {NoStop}%
\bibitem [{\citenamefont {Suma}\ \emph {et~al.}(2018)\citenamefont {Suma},
  \citenamefont {Acun}, \citenamefont {Zorlutuna},\ and\ \citenamefont
  {Vural}}]{suma2018interdependence}%
  \BibitemOpen
  \bibfield  {author} {\bibinfo {author} {\bibfnamefont {D.}~\bibnamefont
  {Suma}}, \bibinfo {author} {\bibfnamefont {A.}~\bibnamefont {Acun}}, \bibinfo
  {author} {\bibfnamefont {P.}~\bibnamefont {Zorlutuna}}, \ and\ \bibinfo
  {author} {\bibfnamefont {D.~C.}\ \bibnamefont {Vural}},\ }\href@noop {}
  {\bibfield  {journal} {\bibinfo  {journal} {Royal Society open science}\
  }\textbf {\bibinfo {volume} {5}},\ \bibinfo {pages} {171395} (\bibinfo {year}
  {2018})}\BibitemShut {NoStop}%
\bibitem [{\citenamefont {Zhao}\ \emph {et~al.}(2016)\citenamefont {Zhao},
  \citenamefont {Liu}, \citenamefont {Zhang}, \citenamefont {Liang},
  \citenamefont {Ding}, \citenamefont {Xu}, \citenamefont {Fang},\ and\
  \citenamefont {Zhang}}]{zhao2016enhanced}%
  \BibitemOpen
  \bibfield  {author} {\bibinfo {author} {\bibfnamefont {L.}~\bibnamefont
  {Zhao}}, \bibinfo {author} {\bibfnamefont {X.}~\bibnamefont {Liu}}, \bibinfo
  {author} {\bibfnamefont {Y.}~\bibnamefont {Zhang}}, \bibinfo {author}
  {\bibfnamefont {X.}~\bibnamefont {Liang}}, \bibinfo {author} {\bibfnamefont
  {Y.}~\bibnamefont {Ding}}, \bibinfo {author} {\bibfnamefont {Y.}~\bibnamefont
  {Xu}}, \bibinfo {author} {\bibfnamefont {Z.}~\bibnamefont {Fang}}, \ and\
  \bibinfo {author} {\bibfnamefont {F.}~\bibnamefont {Zhang}},\ }\href@noop {}
  {\bibfield  {journal} {\bibinfo  {journal} {Experimental cell research}\
  }\textbf {\bibinfo {volume} {344}},\ \bibinfo {pages} {30} (\bibinfo {year}
  {2016})}\BibitemShut {NoStop}%
\bibitem [{\citenamefont {Li}\ \emph {et~al.}(2010)\citenamefont {Li},
  \citenamefont {Zuo}, \citenamefont {He}, \citenamefont {Yang}, \citenamefont
  {Pasha}, \citenamefont {Wang},\ and\ \citenamefont {Xu}}]{li2010paracrine}%
  \BibitemOpen
  \bibfield  {author} {\bibinfo {author} {\bibfnamefont {H.}~\bibnamefont
  {Li}}, \bibinfo {author} {\bibfnamefont {S.}~\bibnamefont {Zuo}}, \bibinfo
  {author} {\bibfnamefont {Z.}~\bibnamefont {He}}, \bibinfo {author}
  {\bibfnamefont {Y.}~\bibnamefont {Yang}}, \bibinfo {author} {\bibfnamefont
  {Z.}~\bibnamefont {Pasha}}, \bibinfo {author} {\bibfnamefont
  {Y.}~\bibnamefont {Wang}}, \ and\ \bibinfo {author} {\bibfnamefont
  {M.}~\bibnamefont {Xu}},\ }\href@noop {} {\bibfield  {journal} {\bibinfo
  {journal} {American Journal of Physiology-Heart and Circulatory Physiology}\
  }\textbf {\bibinfo {volume} {299}},\ \bibinfo {pages} {H1772} (\bibinfo
  {year} {2010})}\BibitemShut {NoStop}%
\bibitem [{\citenamefont {Jiang}\ \emph {et~al.}(2006)\citenamefont {Jiang},
  \citenamefont {Haider}, \citenamefont {Idris}, \citenamefont {Salim},\ and\
  \citenamefont {Ashraf}}]{jiang2006supportive}%
  \BibitemOpen
  \bibfield  {author} {\bibinfo {author} {\bibfnamefont {S.}~\bibnamefont
  {Jiang}}, \bibinfo {author} {\bibfnamefont {H.~K.}\ \bibnamefont {Haider}},
  \bibinfo {author} {\bibfnamefont {N.~M.}\ \bibnamefont {Idris}}, \bibinfo
  {author} {\bibfnamefont {A.}~\bibnamefont {Salim}}, \ and\ \bibinfo {author}
  {\bibfnamefont {M.}~\bibnamefont {Ashraf}},\ }\href@noop {} {\bibfield
  {journal} {\bibinfo  {journal} {Circulation research}\ }\textbf {\bibinfo
  {volume} {99}},\ \bibinfo {pages} {776} (\bibinfo {year} {2006})}\BibitemShut
  {NoStop}%
\bibitem [{\citenamefont {Lin}\ and\ \citenamefont
  {Anseth}(2011)}]{lin2011cell}%
  \BibitemOpen
  \bibfield  {author} {\bibinfo {author} {\bibfnamefont {C.-C.}\ \bibnamefont
  {Lin}}\ and\ \bibinfo {author} {\bibfnamefont {K.~S.}\ \bibnamefont
  {Anseth}},\ }\href@noop {} {\bibfield  {journal} {\bibinfo  {journal}
  {Proceedings of the National Academy of Sciences}\ }\textbf {\bibinfo
  {volume} {108}},\ \bibinfo {pages} {6380} (\bibinfo {year}
  {2011})}\BibitemShut {NoStop}%
\bibitem [{\citenamefont {Mil{\'a}n}\ \emph {et~al.}(2002)\citenamefont
  {Mil{\'a}n}, \citenamefont {P{\'e}rez},\ and\ \citenamefont
  {Cohen}}]{milan2002short}%
  \BibitemOpen
  \bibfield  {author} {\bibinfo {author} {\bibfnamefont {M.}~\bibnamefont
  {Mil{\'a}n}}, \bibinfo {author} {\bibfnamefont {L.}~\bibnamefont
  {P{\'e}rez}}, \ and\ \bibinfo {author} {\bibfnamefont {S.~M.}\ \bibnamefont
  {Cohen}},\ }\href@noop {} {\bibfield  {journal} {\bibinfo  {journal}
  {Developmental cell}\ }\textbf {\bibinfo {volume} {2}},\ \bibinfo {pages}
  {797} (\bibinfo {year} {2002})}\BibitemShut {NoStop}%
\bibitem [{\citenamefont {Boissard}\ \emph {et~al.}(2017)\citenamefont
  {Boissard}, \citenamefont {Tosolini}, \citenamefont {Ligat}, \citenamefont
  {Quillet-Mary}, \citenamefont {Lopez}, \citenamefont {Fourni{\'e}},
  \citenamefont {Ysebaert},\ and\ \citenamefont {Poupot}}]{boissard2017nurse}%
  \BibitemOpen
  \bibfield  {author} {\bibinfo {author} {\bibfnamefont {F.}~\bibnamefont
  {Boissard}}, \bibinfo {author} {\bibfnamefont {M.}~\bibnamefont {Tosolini}},
  \bibinfo {author} {\bibfnamefont {L.}~\bibnamefont {Ligat}}, \bibinfo
  {author} {\bibfnamefont {A.}~\bibnamefont {Quillet-Mary}}, \bibinfo {author}
  {\bibfnamefont {F.}~\bibnamefont {Lopez}}, \bibinfo {author} {\bibfnamefont
  {J.-J.}\ \bibnamefont {Fourni{\'e}}}, \bibinfo {author} {\bibfnamefont
  {L.}~\bibnamefont {Ysebaert}}, \ and\ \bibinfo {author} {\bibfnamefont
  {M.}~\bibnamefont {Poupot}},\ }\href@noop {} {\bibfield  {journal} {\bibinfo
  {journal} {Oncotarget}\ }\textbf {\bibinfo {volume} {8}},\ \bibinfo {pages}
  {52225} (\bibinfo {year} {2017})}\BibitemShut {NoStop}%
\bibitem [{\citenamefont {Davison}\ \emph {et~al.}(2013)\citenamefont
  {Davison}, \citenamefont {Durbin}, \citenamefont {Thau}, \citenamefont
  {Zellmer}, \citenamefont {Chapman}, \citenamefont {Diener}, \citenamefont
  {Wathen}, \citenamefont {Leevy},\ and\ \citenamefont
  {Schafer}}]{davison2013antioxidant}%
  \BibitemOpen
  \bibfield  {author} {\bibinfo {author} {\bibfnamefont {C.~A.}\ \bibnamefont
  {Davison}}, \bibinfo {author} {\bibfnamefont {S.~M.}\ \bibnamefont {Durbin}},
  \bibinfo {author} {\bibfnamefont {M.~R.}\ \bibnamefont {Thau}}, \bibinfo
  {author} {\bibfnamefont {V.~R.}\ \bibnamefont {Zellmer}}, \bibinfo {author}
  {\bibfnamefont {S.~E.}\ \bibnamefont {Chapman}}, \bibinfo {author}
  {\bibfnamefont {J.}~\bibnamefont {Diener}}, \bibinfo {author} {\bibfnamefont
  {C.}~\bibnamefont {Wathen}}, \bibinfo {author} {\bibfnamefont {W.~M.}\
  \bibnamefont {Leevy}}, \ and\ \bibinfo {author} {\bibfnamefont {Z.~T.}\
  \bibnamefont {Schafer}},\ }\href@noop {} {\bibfield  {journal} {\bibinfo
  {journal} {Cancer research}\ }\textbf {\bibinfo {volume} {73}},\ \bibinfo
  {pages} {3704} (\bibinfo {year} {2013})}\BibitemShut {NoStop}%
\bibitem [{\citenamefont {Fonseca}\ \emph {et~al.}(2007)\citenamefont
  {Fonseca}, \citenamefont {Reis}, \citenamefont {Coelho}, \citenamefont
  {Nogueira}, \citenamefont {Monteiro}, \citenamefont {Mairena}, \citenamefont
  {Bacal}, \citenamefont {Bocchi}, \citenamefont {Guilherme}, \citenamefont
  {Zheng} \emph {et~al.}}]{fonseca2007locally}%
  \BibitemOpen
  \bibfield  {author} {\bibinfo {author} {\bibfnamefont {S.}~\bibnamefont
  {Fonseca}}, \bibinfo {author} {\bibfnamefont {M.}~\bibnamefont {Reis}},
  \bibinfo {author} {\bibfnamefont {V.}~\bibnamefont {Coelho}}, \bibinfo
  {author} {\bibfnamefont {L.}~\bibnamefont {Nogueira}}, \bibinfo {author}
  {\bibfnamefont {S.}~\bibnamefont {Monteiro}}, \bibinfo {author}
  {\bibfnamefont {E.}~\bibnamefont {Mairena}}, \bibinfo {author} {\bibfnamefont
  {F.}~\bibnamefont {Bacal}}, \bibinfo {author} {\bibfnamefont
  {E.}~\bibnamefont {Bocchi}}, \bibinfo {author} {\bibfnamefont
  {L.}~\bibnamefont {Guilherme}}, \bibinfo {author} {\bibfnamefont
  {X.}~\bibnamefont {Zheng}},  \emph {et~al.},\ }\href@noop {} {\bibfield
  {journal} {\bibinfo  {journal} {Scandinavian journal of immunology}\ }\textbf
  {\bibinfo {volume} {66}},\ \bibinfo {pages} {362} (\bibinfo {year}
  {2007})}\BibitemShut {NoStop}%
\bibitem [{\citenamefont {Tamm}\ and\ \citenamefont
  {Kikuchi}(1990)}]{tamm1990insulin}%
  \BibitemOpen
  \bibfield  {author} {\bibinfo {author} {\bibfnamefont {I.}~\bibnamefont
  {Tamm}}\ and\ \bibinfo {author} {\bibfnamefont {T.}~\bibnamefont {Kikuchi}},\
  }\href@noop {} {\bibfield  {journal} {\bibinfo  {journal} {Journal of
  cellular physiology}\ }\textbf {\bibinfo {volume} {143}},\ \bibinfo {pages}
  {494} (\bibinfo {year} {1990})}\BibitemShut {NoStop}%
\bibitem [{\citenamefont {Jones}\ \emph {et~al.}(2014)\citenamefont {Jones},
  \citenamefont {Scheuerlein}, \citenamefont {Salguero-G{\'o}mez},
  \citenamefont {Camarda}, \citenamefont {Schaible}, \citenamefont {Casper},
  \citenamefont {Dahlgren}, \citenamefont {Ehrl{\'e}n}, \citenamefont
  {Garc{\'\i}a}, \citenamefont {Menges} \emph {et~al.}}]{jones2014diversity}%
  \BibitemOpen
  \bibfield  {author} {\bibinfo {author} {\bibfnamefont {O.~R.}\ \bibnamefont
  {Jones}}, \bibinfo {author} {\bibfnamefont {A.}~\bibnamefont {Scheuerlein}},
  \bibinfo {author} {\bibfnamefont {R.}~\bibnamefont {Salguero-G{\'o}mez}},
  \bibinfo {author} {\bibfnamefont {C.~G.}\ \bibnamefont {Camarda}}, \bibinfo
  {author} {\bibfnamefont {R.}~\bibnamefont {Schaible}}, \bibinfo {author}
  {\bibfnamefont {B.~B.}\ \bibnamefont {Casper}}, \bibinfo {author}
  {\bibfnamefont {J.~P.}\ \bibnamefont {Dahlgren}}, \bibinfo {author}
  {\bibfnamefont {J.}~\bibnamefont {Ehrl{\'e}n}}, \bibinfo {author}
  {\bibfnamefont {M.~B.}\ \bibnamefont {Garc{\'\i}a}}, \bibinfo {author}
  {\bibfnamefont {E.~S.}\ \bibnamefont {Menges}},  \emph {et~al.},\ }\href@noop
  {} {\bibfield  {journal} {\bibinfo  {journal} {Nature}\ }\textbf {\bibinfo
  {volume} {505}},\ \bibinfo {pages} {169} (\bibinfo {year}
  {2014})}\BibitemShut {NoStop}%
\bibitem [{\citenamefont {Petralia}\ \emph {et~al.}(2014)\citenamefont
  {Petralia}, \citenamefont {Mattson},\ and\ \citenamefont
  {Yao}}]{petralia2014aging}%
  \BibitemOpen
  \bibfield  {author} {\bibinfo {author} {\bibfnamefont {R.~S.}\ \bibnamefont
  {Petralia}}, \bibinfo {author} {\bibfnamefont {M.~P.}\ \bibnamefont
  {Mattson}}, \ and\ \bibinfo {author} {\bibfnamefont {P.~J.}\ \bibnamefont
  {Yao}},\ }\href@noop {} {\bibfield  {journal} {\bibinfo  {journal} {Ageing
  research reviews}\ }\textbf {\bibinfo {volume} {16}},\ \bibinfo {pages} {66}
  (\bibinfo {year} {2014})}\BibitemShut {NoStop}%
\bibitem [{\citenamefont {Yao}\ \emph {et~al.}(2012)\citenamefont {Yao},
  \citenamefont {Li},\ and\ \citenamefont {Ding}}]{yao2012interstitial}%
  \BibitemOpen
  \bibfield  {author} {\bibinfo {author} {\bibfnamefont {W.}~\bibnamefont
  {Yao}}, \bibinfo {author} {\bibfnamefont {Y.}~\bibnamefont {Li}}, \ and\
  \bibinfo {author} {\bibfnamefont {G.}~\bibnamefont {Ding}},\ }\href@noop {}
  {\bibfield  {journal} {\bibinfo  {journal} {Evidence-Based Complementary and
  Alternative Medicine}\ }\textbf {\bibinfo {volume} {2012}} (\bibinfo {year}
  {2012})}\BibitemShut {NoStop}%
\bibitem [{\citenamefont {Swartz}\ and\ \citenamefont
  {Lund}(2012)}]{swartz2012lymphatic}%
  \BibitemOpen
  \bibfield  {author} {\bibinfo {author} {\bibfnamefont {M.~A.}\ \bibnamefont
  {Swartz}}\ and\ \bibinfo {author} {\bibfnamefont {A.~W.}\ \bibnamefont
  {Lund}},\ }\href@noop {} {\bibfield  {journal} {\bibinfo  {journal} {Nature
  Reviews Cancer}\ }\textbf {\bibinfo {volume} {12}},\ \bibinfo {pages} {210}
  (\bibinfo {year} {2012})}\BibitemShut {NoStop}%
\bibitem [{\citenamefont {Rossy}\ \emph {et~al.}(2019)\citenamefont {Rossy},
  \citenamefont {Nadell},\ and\ \citenamefont {Persat}}]{rossy2019cellular}%
  \BibitemOpen
  \bibfield  {author} {\bibinfo {author} {\bibfnamefont {T.}~\bibnamefont
  {Rossy}}, \bibinfo {author} {\bibfnamefont {C.~D.}\ \bibnamefont {Nadell}}, \
  and\ \bibinfo {author} {\bibfnamefont {A.}~\bibnamefont {Persat}},\
  }\href@noop {} {\bibfield  {journal} {\bibinfo  {journal} {Nature
  communications}\ }\textbf {\bibinfo {volume} {10}},\ \bibinfo {pages} {1}
  (\bibinfo {year} {2019})}\BibitemShut {NoStop}%
\end{thebibliography}%


\newpage
\clearpage
\newpage
\clearpage
\renewcommand{\figurename}{Supplementary Figure}
\setcounter{figure}{0}  

\setcounter{equation}{0}
\section*{Appendix}


\section{Additional experimental results}

\begin{figure*}
    \centering
    \includegraphics[width=0.99\textwidth]{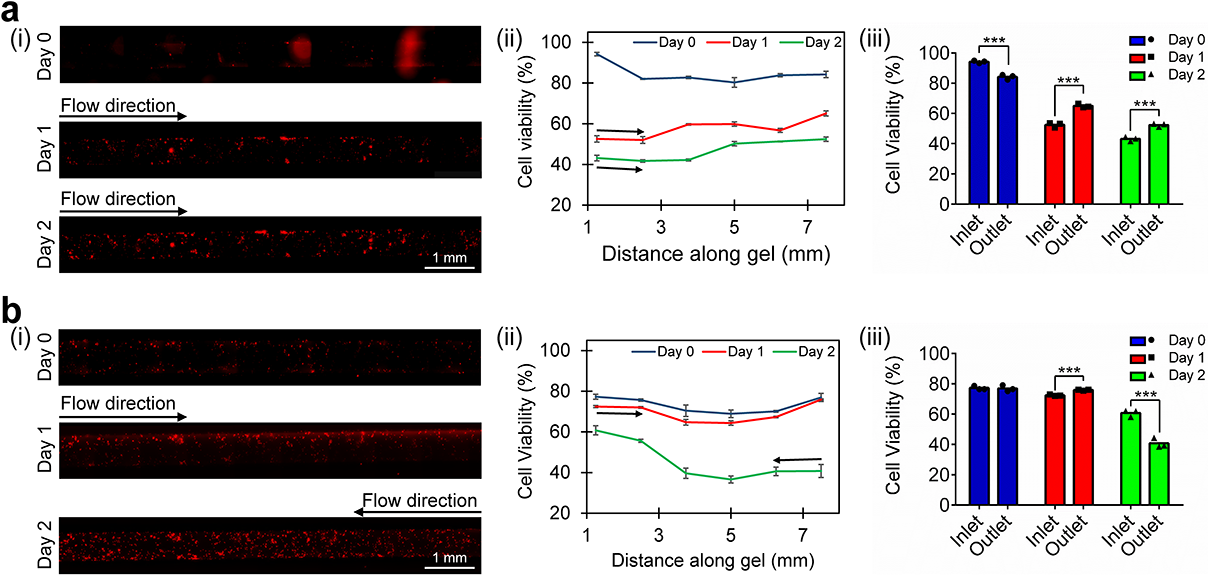}
    \caption{{\bf The change in viability of cardiac fibroblasts along the microchannel upon perfusion.} Cell viability in the PEG:PEG-RGD gel perfused for {\bf (a)} two days from left to right or {\bf (b)} one day from left to right followed by one day from right to left. (i) Fluorescence microscopy images showing the dead cells in the gels at Day 0, Day 1 and Day 2. Red: ethidium homodimer-1. (ii) Line graphs showing cell viability along the gels. (iii) Bar graphs showing cell viability at the inlet and outlet portions of the gels. Flow rate: 20 $\mu$L/min. Asterisks indicate statistical significance $p < 0.05$}
    \label{fig:suppl_exp_one}
\end{figure*}

To see the effect over long time, the engineered tissue was perfused at 20 $\mu$L/h flow rate for 2 days, and the dead cells were imaged and counted at Days 0, 1 and 2 (Supplementary Fig. \ref{fig:suppl_exp_one}a(i)). At Day 0, bright field images were also taken to account for the initial total cell number. Cell viability at Day 0 was similar throughout the gel, except the viability at the inlet (point 1.25 mm) being significantly higher than the other points ($p < 0.0006$ when compared to outlet (point 7.5 mm)) (Supplementary Fig. \ref{fig:suppl_exp_one}a(ii, iii)). Upon perfusion, more cells died at the inlet and cell viability showed an increasing trend along the gel towards the outlet. Viability at the inlet was significantly lower than that at the outlet both for Day 1 ($p < 0.0004$) and Day 2 ($p < 0.0008$). 

In another experiment, we kept all the parameters the same, but reversed the flow direction after Day 1 (Supplementary Fig. \ref{fig:suppl_exp_one}b(i)). At Day 0, cell viability was the same all along the gel. Upon perfusion for 1 day, more cells died at the inlet (left side of the gel) than the outlet (right side), and cell viability followed an increasing trend towards the outlet (Supplementary Fig. \ref{fig:suppl_exp_one}b(ii)). After day 1, flow direction was changed; perfusion was applied from the right side of the gel. After perfusion for another day, but in the reverse direction, we again observed an increasing cell viability towards the outlet (left side of the gel). While, at Day 1, cell viability was significantly higher at the right side of the gel (outlet) than the left ($p < 0.0006$), it was higher at the left side than the right at Day 2 ($p < 0.0009$) (Fig. \ref{fig:suppl_exp_one}a(iii)). 

\section{Chemical Green's function}
\label{sec:greens}
We assume the simplified case of a constant flow profile, $\mathbf{v} = v \hat{z}$. Assuming a constant flow profile, we can use a shift of coordinates to simplify the chemical dynamics. If we let $x_d$ be the coordinate in the direction of flow, and let $z = x_d - vt$, the chemical equation reduces to
\[
\frac{\partial \Phi}{\partial t} = D \nabla^2 \Phi - \gamma \Phi + A n    \label{eq:chemicals_reduced}    
\]
where now, $\boldsymbol{\nabla} = \sum\limits_{i=1}^{d-1} \partial_{x_i} \hat{x}_i + \partial_z \hat{z}$.

Now for an arbitrary source function $S(\mathbf{x},t) = A n(\mathbf{x},t)$, the chemical profile will be given by,
\begin{equation}
    \Phi (\mathbf{x}, t) = G(\mathbf{x},t) * S(\mathbf{x},t) ,
\end{equation}
where $*$ is the convolution operator. The Green's function $G(\mathbf{x},t)$ is the solution to
\[
    \left( \partial_t - D \nabla^2 + \gamma \right) G(\mathbf{x},t) = \delta (\mathbf{x}) \delta (t)      
\]
Taking a Fourier transform with respect to space $(\mathbf{x})$, we get
\[
    \left( \partial_t + k^2 D \nabla^2 + \gamma \right) \tilde{G}(\mathbf{k},t) =  \delta (t) .      
\]
The solution to this is given by,
\[
    \tilde{G} (\mathbf{k},t) = \Theta (t) \exp \left[ - (k^2 D + \gamma) t \right] .
\]
We recognize this as the Green's function for the diffusion operator times a decay factor of $\exp [ -\gamma t]$. We then get, for the Green's function in $d$ dimensions,
\begin{equation}
    G(\mathbf{x},t, v) = \Theta(t) \left( \frac{1}{4 \pi D t} \right)^{d/2} e^{ - r^2 / 4 D t} e^{ - \gamma t}
\end{equation}
where $r^2 = \sum\limits_{i = 1}^{d-1} x_i^2 + (x_d - v t)^2$. 

If we now have a stationary point source, we can get the steady state Green's function by convolving with a source, $S(x,t) = \delta (x)$ constant in time.

In the one dimensional case, we get,
\begin{align*}
    \Phi (x, v) &= \int\limits_{-\infty}^{\infty} \int\limits_{-\infty}^{\infty} G(x' - x,t') \delta (x') \ \mathrm{d}x' \mathrm{d}t' \\
    &= \frac{ \exp \left[ x \left( \frac{v}{2d} - \mathrm{sign} (x) \sqrt{ \gamma / d + v^2 / 4 d^2 } \right) \right]}{2 d \sqrt{ \gamma / d + v^2 / 4 d^2 }} \\
    &\equiv \exp \left[ x \left( \frac{v}{2d} - \mathrm{sign} (x) \lambda \right) \right]/(2 d \lambda)
\end{align*}
where we have defined $\lambda(v) = \sqrt{ \gamma / d + v^2 / 4 d^2 }$. From here, we can get diffusion-advection-decay lengths given by
\begin{equation}
    \ell_{L} (v) = \left[ \lambda(v) + \frac{v}{2d} \right]^{-1} \quad
    \ell_{R} (v) = \left[ \lambda(v) - \frac{v}{2d}  \right]^{-1},
    \label{eq:suppl_leftright}
\end{equation}
assuming the flow is from left to right $(v \geq 0)$.

\section{Failure progagation velocity derivation}
\label{sec:derivations}
We begin by assuming that the cells respond to the chemical concentration ``sharply''. This is the case as $k \rightarrow \infty$, where the response function becomes a step function. We then later relax this assumption. 

To simplify our analysis, we used a boxcar approximation to the Green's function for the cooperative factors where left and right lengths are given by eqn.\ref{eq:suppl_leftright}, and assume chemicals quickly reach steady state. The total area of the Green's function corresponds to the secretion rate per cell density at steady state. We therefore have 
\begin{equation}
    \label{eq:suppl_boxcar_green}
    G(x) = (A/\gamma) \Theta(x + \ell_{L} ) \Theta( \ell_{R} -x ),
\end{equation}
where $\Theta(x)$ is the Heaviside step function. 

To get a first approximation to the cooperative factor concentration, we convolve the boxcar Green's function (equation \ref{eq:suppl_boxcar_green}) with an initial semi-infinite concentration of cells, $n(x,0) = n_0 \Theta(x)$. This gives us a linearly increasing chemical profile, with slope $H/W$, that saturates to a maximum value of $H$ after a length of $\ell_{R}$ into the microchannel, where $H = A n_0 / \gamma$ and $W = \ell_{L} + \ell_{R}$. Specifically, we get
\begin{equation}
    \Phi(x) = 
    \begin{cases} 
        0, & x < -\ell_{L} \\ 
        \frac{H}{W} \left(x + \ell_{L} \right), &  -\ell_{L} < x < \ell_{R}  \\
         H, & x > \ell_{R} 
    \end{cases}
\end{equation}

Now, if the response to the chemicals is a step function (as is the case when $k \rightarrow \infty$), and if $H \gg \phi_0$, cells where $\Phi(x)$ is below $\phi_0$ will die at a rate $\alpha$ and cells where $\Phi(x)$ is above $\phi_0$ will survive. The cell density will then shift to the right by an amount $\Delta$ given by setting $\Phi(\Delta) = \phi_0$ or, 
\begin{equation}
    \Delta = \frac{1}{2 A \gamma n_0} \left[ A n_0 v + \left(2 \gamma \phi_0 - A n_0 \right) \sqrt{4 d \gamma + v^2} \right]. \label{eq:suppl_simple_vel}
\end{equation}
The cooperative factor concentration will then update with this new concentration of cells and we can then reiterate this to get the next death of cells. The death of cells will then continue to propagate by an amount $\Delta$ at a rate $\alpha$. The death propagation velocity is therefore given as $v_{d} = \alpha \Delta$. We find this gives good agreement for large $k$ and short times, but fails to capture the strong time dependence for small $k$.

We then improve on our analytical results by considering the death of the bulk (in the region $x> \ell_{R}$). We replace the initial cell concentration $n_0$ in equation (\ref{eq:suppl_simple_vel}) by a time varying function $n_b (t)$ for the bulk cell concentration. To get $n_b (t)$, we approximate the chemical concentration in the bulk as the steady state, non-spatial concentration, $\Phi^* = A n_b(t) /\gamma$. We then expand the hill form about infinity (high concentration limit), to get 
\[
    \frac{1}{ (\Phi/\phi_0)^k + 1} \approx \left( \frac{\phi_0}{\Phi} \right)^k - \left( \frac{\phi_0}{\Phi} \right)^{2k} + \left( \frac{\phi_0}{\Phi} \right)^{3k} -\ldots
\]
Taking the first order term and plugging in $\Phi^*$, we get for the attenuation of the bulk cell density,
\begin{equation}
    n_b(t) = \left[ n_0^k - \alpha \left(\frac{\gamma \phi_0}{A} \right)^k k t \right]^{1/k}.
    \label{eq:suppl_app_bulk}
\end{equation}
Plugging this into the above expression for $v_{d} = \alpha \Delta$, we get,
\begin{equation}
    v_{d} \!=\! \frac{\alpha v}{2 \gamma} \!+\! \frac{\alpha \sqrt{4 d \gamma + v^2}}{\gamma} \left\{\left[ \left(\frac{A n_0}{\gamma \phi_0}\right)^k \!\!\!-\! \alpha k t \right]^{-\frac{1}{k}} \!\!\!\!-\! \frac{1}{2} \right\}.\nonumber 
\end{equation}
This expression for the velocity better captures the acceleration of velocity over time, but grows in error over time for small $k$. This is because we fail to consider the region where $\Phi > \phi_0$ but not yet fully saturated. This gives another region where $n(x,t)$ does not die out exponentially, but dies out faster than $n_b (t)$ and will lead to an increase in the velocity $v_d$.

Therefore, for a better approximation to the velocity, we also take into account the region where the cooperative factor concentration is larger than the threshold $\phi_0$, but not yet saturated. The cell concentration in this region (denoted $n_\delta (x,t)$) will decay faster than the cell concentration in the saturated region, $n_b (t)$. 

\begin{figure}
    \centering
    \includegraphics[width=0.49\textwidth]{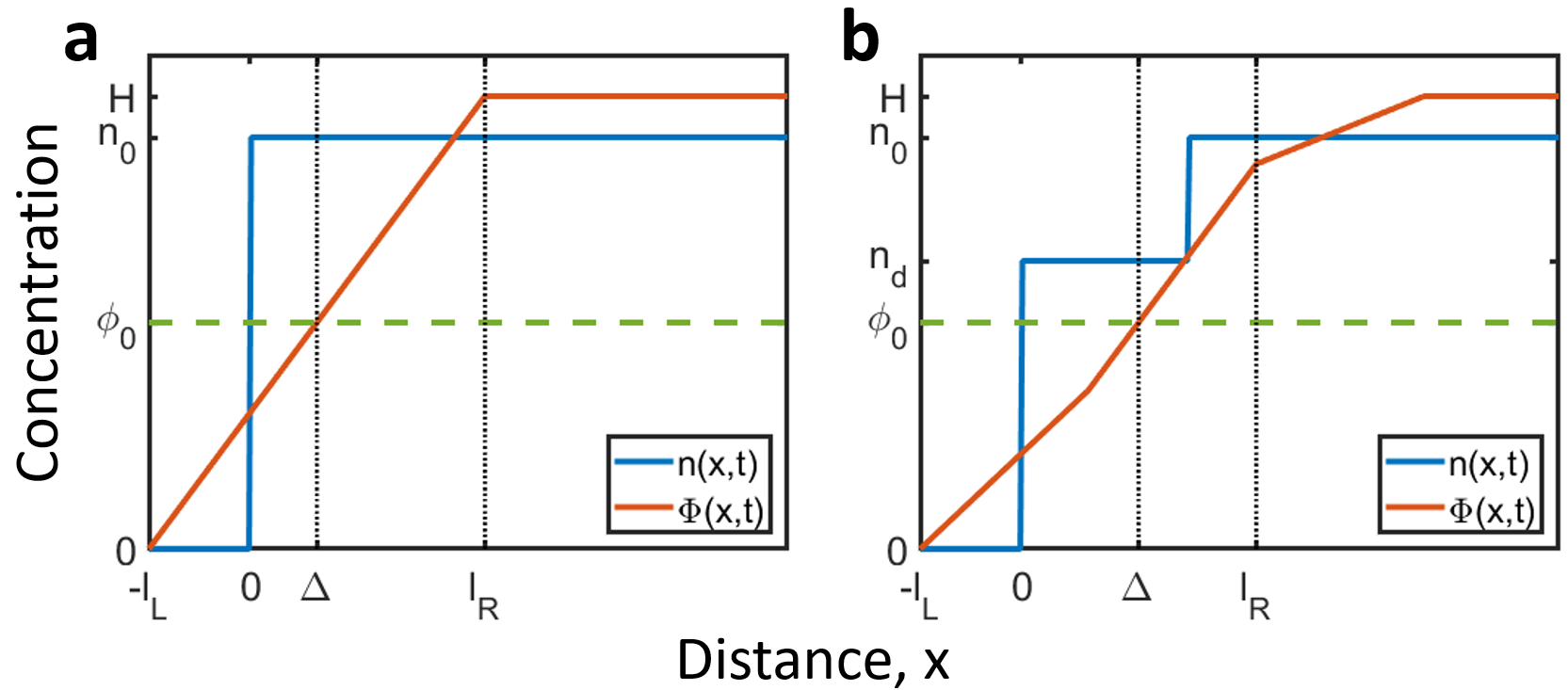}
    \caption{{\bf Schematic of theoretical methods.} {\bf (a)} First iteration. The initial semi-infinite concentration of cells results in a piecewise linear concentration of the cooperative factors. The position where the cooperative factors equal the threshold value $\phi_0$ is given by $\Delta$. Below this value, cells will die out exponentially. A first approximation to the failure propagation velocity can then be given by $v_d = \alpha \Delta$. {\bf (b)} Second iteration. To improve our calculation of the failure propagation velocity, we account for the fact that the region between $\Delta$ and $\ell_R$ in the first iteration dies out faster than the region after $\ell_R$. We approximate this intermediate region by a flat spatially constant value that is less than the cell concentration after $\ell_R$. We then solve again for when the cooperative factor concentration equals the threshold value $\phi_0$ and get a larger value for $\Delta$.}
    \label{fig:thy_schematic}
\end{figure}

We then approximate the cell concentration $n(x,t)$ as $n_\delta (t)$ in the region where $ \phi_0 < \Phi(x) < A n_b / \gamma $ and $n_b(t)$ where $\Phi(x) > A n_b / \gamma$, as shown in Fig. \ref{fig:thy_schematic}. 

Taking the convolution of this updated cell concentration with our boxcar Green's function, we get another piecewise function for $\Phi(x)$, with 5 regions. Our second iteration of $\Phi(x)$ then gives,
\begin{align}
    \Phi = 
    \begin{cases} 
        0 & x < - \ell_L \\ 
        \frac{H_\delta}{W} (x + \ell_L) & -\ell_L < x < -\ell_L + \delta \\ 
        \frac{H_\infty}{W} (x + \ell_L) - \frac{\delta (H_\infty - H_\delta)}{W} & -\ell_L + \delta < x < \ell_R \\
        H_\delta + \frac{H_\infty - H_\delta}{W} (x + \ell_L - \delta) & \ell_R < x < \ell_R + \delta \\
        H_\infty & x > \ell_R + \delta
    \end{cases}\nonumber
\end{align}
where $H_\delta = A n_\delta /\gamma$ and $H_\infty = A n_b /\gamma$. We now solve again for $\Phi(\Delta) = \phi_0$ to get a new expression for $\Delta$. We assume $\delta$ is small, and use the center condition to solve for the threshold crossing. This assumption works for large velocities as $\Delta \rightarrow \ell_R$ and $\delta = \ell_R - \Delta \rightarrow 0$. Solving for the threshold crossing then gives, $\Delta = (W \phi_0 - H_\infty (\ell_L - \ell_R) - H_\delta \ell_R )/( 2 H_\infty - H_\delta)$, which in terms of original paramters gives,
\begin{align}
 \Delta = \frac{A v (2 n_b - n_\delta )  + \sqrt{v^2 + 4 d \gamma}( 2 \gamma \phi_0 - A n_\delta ) }{2 A \gamma (2 n_0- n_\delta)},
\end{align}
Note, that this reduces to our original formula if we set $n_\delta = n_b$ as expected.

Now, to get the velocity propagation, we need an expression for $n_\delta$. We could in principle obtain this from solving for $n(x,t)$ with a linear approximation for $\Phi(x)$. We instead use a simplified expression from the following arguments. As the hill constant $k \rightarrow \infty$, the cell concentration will remain constant for any value of $\Phi$ above the threshold $\phi_0$, and the two regions should remain the same, $n_\delta(t) = n_b(t)$. For $k=1$ we approximate $n_\delta \approx \frac{1}{2} n_b$. We therefore approximate $n_\delta (x,t)$ as 
\begin{equation}
\label{eq:suppl_ndelta}
    n_\delta(t) = \left[1 - \left(\frac{1}{2}\right)^k \right] n_b (t).
\end{equation}
Now taking the velocity to be $v = \alpha \Delta$ and substituting equations (\ref{eq:suppl_app_bulk}) and (\ref{eq:suppl_ndelta}) for $n_b$ and $n_\delta$, we get,
\begin{align}
 v_d &= \frac{\alpha v}{2 \gamma} + \frac{\alpha \sqrt{v^2 + 4 d \gamma}}{2(1 + 2^k)} \times \nonumber \\ &\left\{ \frac{1 - 2^k}{\gamma} \!+\!  \frac{2^{1+k} \phi_0}{A} \left[n_0^k - \alpha k t \left(\frac{ \gamma \phi_0}{A}\right)^k \right]^{-\frac{1}{k}} \right\}
 \label{eq:app_finalvel}
\end{align}
We find this gives good agreement with numerical simulations of the model.

We can also integrate this to get the failure penetration depth over time, $x_d (t) = \int_0^t v_d (t') \ \mathrm{d}t'$, giving,
\begin{align}
    x_d  = \frac{\sqrt{v^2 + 4 d \gamma}}{2 (1+ 2^k) A \gamma} \left\{ A t \left( \frac{(1+2^k)v}{ \sqrt{v^2 + 4 d \gamma}} \!-\! (2^k -1) \right)\right.\nonumber \\
    + \left.\frac{ 2^{1+k} \gamma \phi_0 }{\alpha(1-k)} \left( \frac{\gamma \phi_0}{A} \right)^{-k} \! \left[ \left( n_0^k \!-\! \alpha k t \left( \frac{\gamma \phi_0}{A} \right)^k \right)^{\frac{k\!-\!1}{k}} \!-\! n_0^{k\!-\!1} \right]  \right\}\nonumber
\end{align}
for $k \neq 1$, and
\[
    x_d = \frac{v t}{2 \gamma} - \frac{\sqrt{v^2 + 4 d \gamma}}{6 \alpha \gamma}\left\{ \alpha t + 4 \log\left[1 - \alpha t \left(\frac{\gamma \phi_0}{A n_0} \right) \right] \right\}\nonumber
\]
for $k = 1$.
Also, taking the derivative of equation (\ref{eq:app_finalvel}) gives the acceleration,
\[
    a_d = \frac{2^k \alpha \phi_0 \sqrt{v^2 + 4 d \gamma} \left( \frac{\gamma \phi_0}{A} \right)^k \left[ n_0^k - \alpha k t \left( \frac{\gamma \phi_0}{A} \right)^k \right]^{ -\frac{k+1}{k} } }{(1 + 2^k)A}.\nonumber
\]
\end{document}